\documentclass[aps,prl,twocolumn,showpacs,nofootinbib,superscriptaddress,groupedaddress]{revtex4-1} 
\pdfoutput=1

\usepackage{multirow}
\usepackage{subfigure}
\usepackage{graphicx}
\usepackage{amsfonts}
\usepackage{amsmath,amssymb}
\usepackage{slashed}
\usepackage{feynmp}
\usepackage{caption}
\usepackage{xcolor}
\usepackage{color}
\usepackage[citebordercolor=green,linkbordercolor=red,urlbordercolor=white]{hyperref}
\usepackage{url}
\usepackage{comment}
\usepackage{cancel}
\usepackage{soul}

\begin{document}

\title{Collapsed Dark Matter Structures}

\author{Matthew R.~Buckley}
\affiliation{Department of Physics and Astronomy, Rutgers University, Piscataway, NJ 08854, USA}
\author{Anthony DiFranzo}
\affiliation{Department of Physics and Astronomy, Rutgers University, Piscataway, NJ 08854, USA}

\date{\today}

\begin{abstract}
The distributions of dark matter and baryons in the Universe are known to be very different: the dark matter resides in extended halos, while a significant fraction of the baryons have radiated away much of their initial energy and fallen deep into the potential wells. This difference in morphology leads to the widely held conclusion that dark matter cannot cool and collapse on any scale. We revisit this assumption, and show that a simple model where dark matter is charged under a ``dark electromagnetism'' can allow dark matter to form gravitationally collapsed objects with characteristic mass scales much smaller than that of a Milky Way-type galaxy. Though the majority of the dark matter in spiral galaxies would remain in the halo, such a model opens the possibility that galaxies and their associated dark matter play host to a significant number of collapsed substructures. The observational signatures of such structures are not well explored, but potentially interesting.
\end{abstract}

\maketitle

Though dark matter outmasses the baryons five-to-one \cite{Ade:2015xua}, we typically assume the baryonic components are far more complex than their dark matter equivalents. While some of this is certainly due to baryonic chauvinism, it is clear from rotation curves and lensing measurements that, at the mass scales of dwarf galaxies $({\sim}10^{9}M_\odot)$ and above, dark matter resides in approximately spherical ``halos'' whose shapes are consistent with primordial overdensities evolving solely under gravity. The baryons on the other hand can form objects with a wide variety of shapes (including the disks in Milky Way (MW)-type spiral galaxies). This non-trivial structure continues down to smaller scales (stars and planets). 

These collapsed baryonic structures are made possible through the existence of a cooling mechanism: the electrons (and the coupled protons) can bleed away the protons' initial kinetic and potential energy and sink deep into the gravitational well. For baryons, the primary cooling mechanism (prior to the formation of the first stars) is collisional excitation of hydrogen by electrons \cite{Thoul:1994ir}. Collapsed dark matter structures have been hypothesized, but only if they are created in the early Universe (e.g.~primordial black holes \cite{russiansPBH,Hawking:1974rv,Ivanov:1994pa} or axion stars \cite{Tkachev:1991ka}), or in some subdominant component of the dark matter \cite{Fan:2013tia,Fan:2013yva,Agrawal:2017rvu,Fischler:2014jda,Mohapatra:1996yy}. Even models of self-interacting dark matter, which can alter the morphology of dark matter halos \cite{Tulin:2017ara}, do not allow for the {\em loss} of energy, but only the {\em transfer} of energy between dark matter particles. In general, we assume all of the dark matter cannot possess the necessary interactions to dissipate energy, because such interactions would seem to violate the known shape of dark matter halos at the scale of dwarf galaxies and larger, which are largely consistent with the predictions of non-interacting, non-cooling, cold dark matter (CDM) \cite{Viel:2005qj,Seljak:2006bg}.

However, this is not the case: cooling mechanisms generically operate efficiently only {\em within} a relatively narrow range of halo masses set by the parameters of the dark sector. Inside this mass range, primordial dark matter halos can cool and collapse into compact objects, fragmenting down to smaller mass objects as they do. Outside it, the characteristic cooling time is longer than the infall time. As a result, little of the kinetic and potential energy in the halo is lost. Thus, it is completely possible for compact objects to form at the scale of, say $10^6M_\odot$ and below, while above this scale, no significant deviation from CDM would be seen, {\em despite} 100\% of the dark matter having the same set of non-gravitational interactions.

This proposed cooling mechanism is in exact analogy to the excitational cooling of bound hydrogen by electrons, which sets the characteristic mass of the largest collapsed baryonic structures in the Universe to be ${\sim}10^{12}M_\odot$ \cite{Silk:1977wz} (see also \cite{White:1977jf,Silk:2013xca}). The key observation is that this cooling rate goes as $T_{V}^{-1/2}$, while the energy of the particles increases as the virial temperature $T_{V}$. As halo mass (and thus $T_{V}$) increases, cooling becomes proportionally less efficient, setting an upper limit on the mass of halos that could cool and collapse. This is why collections of collapsed, fragmented baryons exist on mass scales below $10^{12}M_\odot$, while above this scale the galaxy clusters are roughly spherical collections of virialized baryonic gas in which the galaxies are embedded -- the baryons in the galaxy cluster have too much kinetic energy and the cooling mechanism cannot remove an ${\cal O}(1)$ fraction within the characteristic free-fall time. Going to smaller halo masses, the ionization fraction decreases with the virial temperature, as does the scattering rate of free electrons off of bound hydrogen. This sets a lower limit on the mass of collapsed objects.\footnote{This assumes no additional source of reionization is present. For baryons, stars will eventually provide ionizing photons, allowing the baryons in small halos to collapse, but we will not appeal to a similar mechanism in the dark sector.}

The range of masses of baryon halos which can collapse is set by the masses and couplings of protons and electrons; we can hypothesize that the dark sector could have a similar mechanism with different parameters, leading to a different mass range that does not extend up to the scale of spiral galaxies. As a proof of principle, we introduce a simple model, where dark matter is composed of equal numbers of a heavy particle $H$ and a light particle $L$, with opposite charges under an unbroken $U(1)_X$ with fine-structure constant $\alpha_D$ (a ``dark electromagnetism'' \cite{Ackerman:mha}). Such dark matter is asymmetric (see \cite{Zurek:2013wia} for a review), and so must have additional non-trivial interactions in the early Universe to annihilate away the thermal component \cite{Buckley:2011kk}, which we will take as given. Models of dark matter charged under a hidden $U(1)$ have been considered in the literature, e.g.~\cite{Feng:2009mn,Kaplan:2009de,CyrRacine:2012fz}. In this Letter, our goal is to show that this simple model contains the necessary components to allow the formation of collapsed structures in the dark sector and evades other constraints. We note that dissipative cooling of dark matter has been considered previously \cite{Foot:2014mia,Foot:2014uba,Boddy:2016bbu,Rosenberg:2017qia}, and for a subdominant component of dark matter in \cite{Fan:2013tia,Fan:2013yva,Cyr-Racine:2013fsa,Agrawal:2017pnb,Agrawal:2017rvu}, but we believe this is the first work that emphasizes that the cooling mechanism selects a critical mass scale above which cooling is inefficient.

Our model has three free parameters: the coupling $\alpha_D$ and two masses, $m_H$ and $m_L$. We are most interested in the parameter space $m_H \gg m_p$, in order to avoid constraints from elastic scattering. Assuming the visible and dark sectors were once in thermal equilibrium, the photon bath and dark sector temperatures today differ according to the ratio $\xi = T_D/T_\gamma$ set by the effective number of relativistic degrees of freedom today as compared to when the dark and visible sectors decoupled from each other \cite{Ackerman:mha,Agrawal:2017rvu}:
\begin{equation}
\xi = \left[\frac{g^{\rm vis}_{*s}(T)}{g^{\rm vis}_{*s}(T_{\rm dec})}  \frac{g^{\rm DM}_{*s}(T_{\rm dec})}{g^{\rm DM}_{*s}(T)}\right]^{1/3}
\end{equation}
If both $H$ and $L$ are relativistic when the sectors decoupled and there are no additional degrees of freedom in the visible sector, then $\xi \sim 0.5$, dropping to ${\sim} 0.4$ if $H$ is already non-relativistic at decoupling. This continues to decrease if more particles are added to the visible sector. A more complicated thermal history could further reduce $\xi$, for example, if the reheating mechanism favors one sector over the other \cite{Berezhiani:1995am,Hall:2009bx,Allahverdi:2010xz,Adshead:2016xxj}. As it depends on unspecified high energy physics which are beyond the scope of this Letter, we treat $\xi$ as a free parameter, noting that $\xi \sim 0.2-0.5$ are typical values for minimal dark matter models with a standard thermal history.

In the early Universe, nearly scale-free primordial density fluctuations are seeded by inflation \cite{Bardeen:1983qw}. These overdensities will increase in size due to gravitational collapse, departing from linear growth when the fractional overdensity is $\delta \sim 1.7$, and virializing at $\delta_V \sim 178$ \cite{Blumenthal:1984bp}. For a halo of total mass $M$, the virial temperature of the $H$ particles will be:
\begin{equation}
T_V \sim \left( \frac{4\pi}{3}\delta_V \bar{\rho}(z) \right)^{1/3} G_N m_H M^{2/3},
\end{equation}
where $\bar{\rho}(z)$ is the average density of dark matter at redshift $z$. For the mass range of interest (${\lesssim} 10^9~M_\odot$), we expect the dark matter halos to be virialized by $z \sim 10$ \cite{Springel:2005nw,Behroozi:2014tna}, which we adopt as our benchmark. The lighter $L$ particles are kept in equilibrium with the $H$ via Coulomb scattering \cite{Agrawal:2017rvu}; the timescale of which is much shorter than other processes, so we treat both particles as having a common virial temperature $T_V$. 
 
In the Standard Model, there are four primary mechanisms for the baryons to lose energy: inverse Compton scattering of Cosmic Microwave Background (CMB) photons, bremsstrahlung, free-bound scattering where the electrons bind to a proton, and collisional excitation of hydrogen by a free electron. As the temperature of the ``dark CMB'' must be lower than our own to evade constraints on the number of light degrees of freedom in the early Universe \cite{Feng:2008mu}, we treat ``dark inverse Compton'' as subdominant for the purposes of our proof of principle. Of the remaining cooling mechanisms, the most important for us (and for the earliest baryonic halos) is collisional excitation from the $1s\to 2p$ state, which has a volumetric cooling rate of \cite{Rosenberg:2017qia}:
\begin{eqnarray}
\Pi &=& \frac{2^{16}}{3^9}\sqrt{\frac{2\pi}{m_L T_V}}\alpha_D^2 n_L(z) n_B(z)\\
& \times & \int_{\sqrt{3}y/2}^\infty du \frac{ue^{-u^2}}{1+\frac{7y^2}{4u^2}} \int_{x_-}^{x_+}\frac{dx}{x}\left(1+\frac{4x^2}{9} \right)^{-6}\nonumber,
\end{eqnarray}
where $y = \sqrt{\alpha_D^2m_L/2T_V}$, $x_\pm = \frac{u}{y}\left(1\pm\sqrt{1-3y^2/4u^2}\right)$, and $n_L(z)$ ($n_B(z)$) is the number density of free $L$ (bound states) at redshift $z$. Clearly, for this cooling to be effective, there must be a non-negligible number density of both bound and free states, which occurs over a relatively narrow temperature range. The ionization fraction is set by the relative rates of collisional ionization and recombination: $n_L/n_B = \langle \sigma_{\rm ci} v\rangle/\langle \sigma_{\rm rec} v\rangle$ \cite{Mo:2010gfe}. Using the binary-encounter Bethe model \cite{Rosenberg:2017qia,PhysRevA.50.3954} and setting the bound-free Gaunt factor \cite{Karzas:1961ApJS} to one:
\begin{eqnarray}
\langle \sigma_{\rm ci} v\rangle & = & \sqrt{\frac{2^7 \pi}{m_L^3 T_V}} \int_y^\infty du\frac{ue^{-u^2}}{1+\tfrac{2y^2}{u^2}}\times \\
&&\left[1-\tfrac{y^2}{u^2}-\tfrac{1}{2}\left(1- \tfrac{y^4}{u^4} \right)\log\tfrac{y^2}{u^2}+\tfrac{y^2\log\tfrac{y^2}{u^2}}{u^2+y^2} \right], \nonumber \\
\langle \sigma_{\rm rec} v\rangle & = & \alpha_D^5 \sqrt{\frac{2^{11}\pi}{3^3m_L T_V^3}} \int_0^\infty du \sum_{n=1}^\infty \frac{ue^{-u^2}}{u^2n^3+y^2n}.
\end{eqnarray}

The kinetic energy density of the dark matter is $\tfrac{3}{2}n_{\rm DM}T_V$ (where $n_{\rm DM}$ is the total dark matter number density). The characteristic time for the dark matter to radiate away ${\cal O}(1)$ of its kinetic and potential energy is therefore $t_c = \frac{3}{2}n_{\rm DM}T_V/\Pi$. This must be compared to the free-fall time for the halo, which for an isothermal sphere is $t_f = \sqrt{3\pi/32G\delta_V \bar\rho(z)}$. Thus our collapse condition is:
\begin{equation}
 \sqrt{3\pi/32G\delta_V \bar\rho(z)} > \frac{3}{2}n_{\rm DM}T_V/\Pi, \label{eq:collapset}
\end{equation}
which depends on the halo mass $M$ via $T_V$. Note that our assumptions are conservative, as we assume the halo density is constant and ignore the feedback effect of increasing densities as cooling begins. We therefore expect a larger region of parameter space would allow for cooling, but definitive answers to these questions likely require detailed simulation.

Given free rein to chose $\alpha_D$, $m_H$, and $m_L$, the range of collapsing halo masses could be dialed anywhere from galaxy-cluster scale down to sub-Earth masses. However, these parameters are constrained by other measures of dark matter phenomenology.

Charged dark matter is subject to self-scattering constraints. For charged dark matter composed of only one species, the tightest constraint from self-scattering is \cite{Ackerman:mha,Agrawal:2016quu}:
\begin{equation}
\frac{\alpha_D^2}{m_X^3} \lesssim 10^{-11}~\mbox{GeV}^{-3}. \label{eq:scattering}
\end{equation}
As the $L$ particles are heated to the temperature of the $H$, hard scatterings must transfer the equivalent of the $H$'s kinetic energy in order to significantly alter the halo. Thus, we can apply the scattering limit with $m_X = m_H$. 

In addition, we must consider the self-consistency of the model. By adding a massless dark photon, we not only introduce new relativistic degrees of freedom in the early Universe which are constrained by CMB measurements, we also couple dark matter to a relativistic fluid in the early Universe. This will introduce ``dark acoustic oscillations'' (DAO) \cite{CyrRacine:2012fz}, in analogy to baryon acoustic oscillations. Primordial perturbations, which cross inside the sound horizon before dark photon decoupling, will be suppressed by energy transfer mediated by the dark photons and can erase overdensities which would otherwise later collapse. This ``DAO scale'' serves a conservative lower limit, as the true cut-off is the Silk damping scale \cite{1968ApJ...151..459S} which is lower than the DAO scale \cite{Buckley:2014hja}. The dark radiation will decouple from the dark matter at the following redshift \cite{Cyr-Racine:2013fsa,Agrawal:2017rvu}:
\begin{equation}
\!\!z_{\rm dec}\! \approx\! 
\begin{cases}
    8.5 \times 10^8\! \left(\frac{\alpha_D}{0.1}\right)^2\!\left(\frac{m_L}{{\rm GeV}}\right)\!\left(\frac{0.5}{\xi}\right),\!\!\! & \beta \ge 1 \\
    3.0 \times 10^4\! \left(\frac{0.1}{\alpha_D}\right)\!\left(\frac{m_L}{{\rm GeV}}\right)\!\sqrt{\frac{m_H}{{\rm GeV}}}\left(\frac{0.5}{\xi}\right)^2,\!\!\! & \beta < 1
\end{cases}
\label{eq:zdec}
\end{equation}
where $\beta \equiv \frac{10^{16}}{3} \alpha_D^6\xi^2\left( \frac{{\rm GeV}}{m_H}\right)$. In the first case above, the dark radiation decouples due to $H$--$L$ recombination.  In the latter case, dark radiation decouples due to Hubble expansion before $H$ and $L$ can recombine. Assuming $z_{\rm dec}$ is much earlier than matter-radiation equality  ($z_{\rm eq} \sim 3500$), the smallest dark matter halo mass today would be:
\begin{eqnarray}
M_{\rm min} &=& \frac{4\pi}{3} \rho_{\rm DM} r_{\rm DAO}^3,\\
r_{\rm DAO} & = & \frac{1}{\sqrt{3}} \int_0^{(1+z_{\rm dec})^{-1}} \frac{da}{a^2 H(a)} \left[1+\frac{3a\Omega_{DM}}{4\xi^4\Omega_\gamma} \right]^{-1/2}. \nonumber
\end{eqnarray}
The ratio in brackets is the sound speed in the dark fluid. Consistency requires that $M_{\rm min}$ is smaller than the heaviest dark matter halos which satisfy Eq.~\eqref{eq:collapset}, otherwise no surviving halos can cool. From Eq.~\eqref{eq:zdec}, we see $M_{\rm min}$ will also strongly depend on the temperature ratio $\xi$.

\begin{figure*}[th]
\includegraphics[width=\columnwidth]{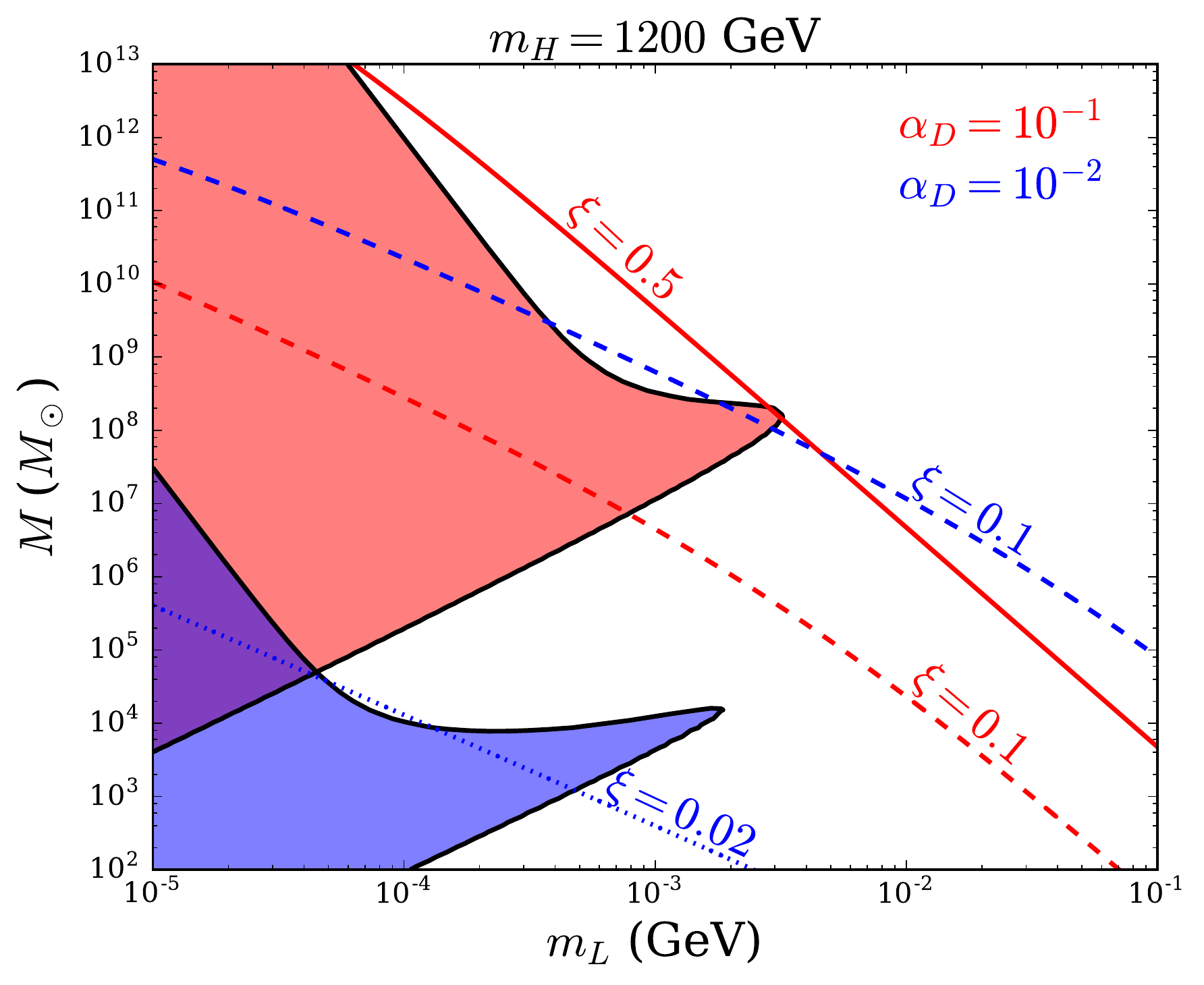}\includegraphics[width=\columnwidth]{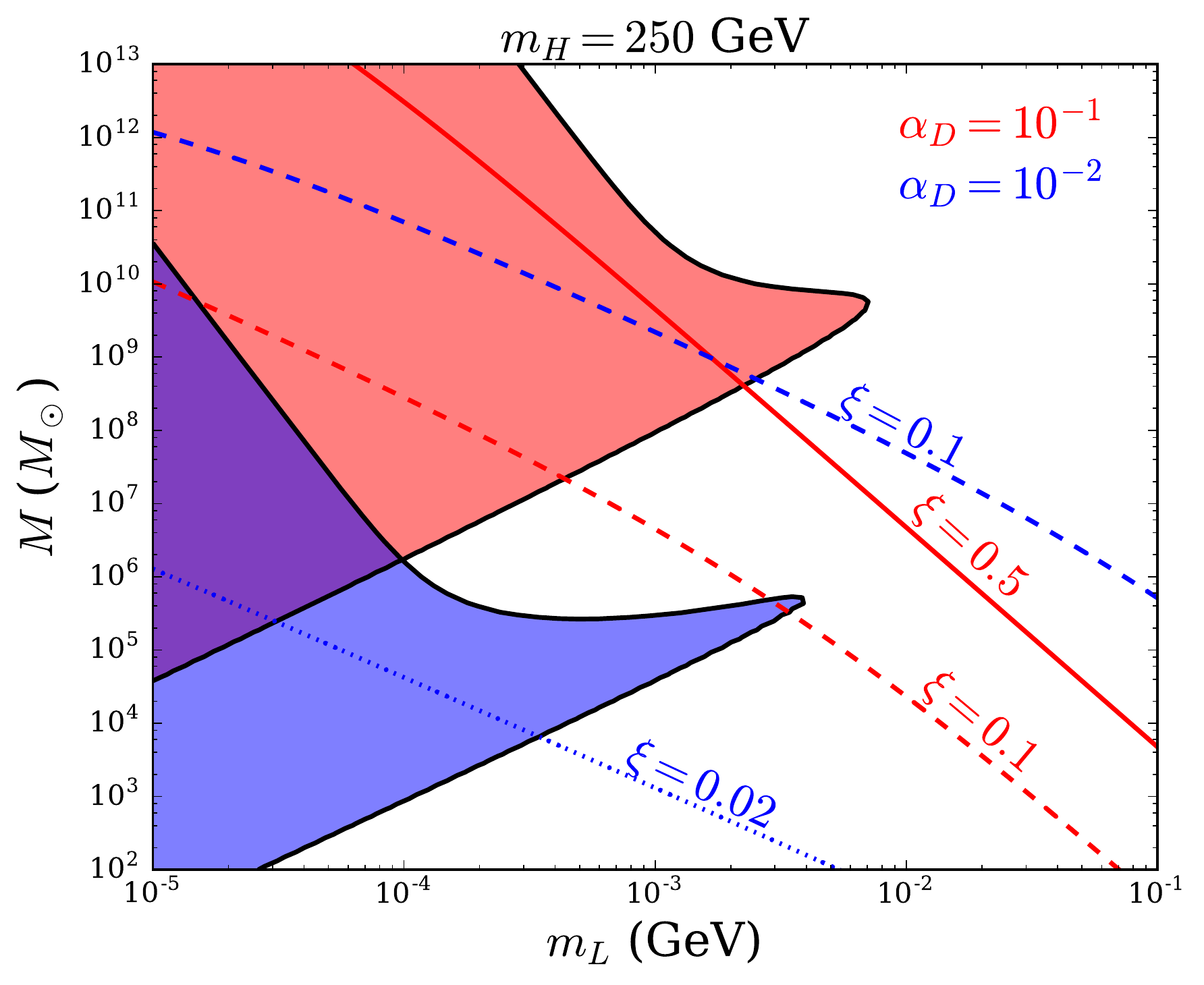}
\caption{Range of dark matter halo masses that cool as a function of $m_L$ for $m_H = 1.2$~TeV (left) and $250$~GeV (right). Colored regions result in halos cooling for $\alpha_D = 10^{-1}$ (red) and $\alpha_D = 10^{-2}$ (blue). Lower limits on the halo mass as a function of $\xi$ are shown as solid, dashed, and dotted lines for $\xi = 0.5$, $0.1$, and $0.02$, respectively (red for $\alpha_D = 10^{-1}$, blue for $\alpha_D =10^{-2}$). For $m_H = 250$~GeV, the $\alpha_D = 0.1$ region violates the self-scattering bound Eq.~\eqref{eq:scattering}, but is included to demonstrate the cooling dependence on $m_H$ and $\alpha_D$.}
\label{fig:mc}
\end{figure*}

In Figure~\ref{fig:mc}, we show the regions of dark matter halo mass $M$ which can cool and collapse as a function of $m_L$, for two values of $m_H$, 1.2~TeV and 250~GeV, and two values of $\alpha_D$, $10^{-1}$ and $10^{-2}$. All cooling mechanisms save Compton scattering are included. Collisional excitation dominates for larger $m_L$, whereas bremsstrahlung becomes important for lower $m_L$, resulting in a kink in the upper range of collapsing halo mass. For $m_H = 250$~GeV, the $\alpha_D = 10^{-1}$ parameter space is excluded by the scattering limit, but all other parameter points are allowed. As can be seen, increasing $m_H$ decreases the critical mass, as does decreasing $\alpha_D$. For our $m_H = 1.2$~TeV, $\alpha_D = 10^{-1}$ working point, $\xi = 0.5$ allows for minimum halo masses at the collapse scale, even with our conservative use of the DAO scale for $M_{\rm min}$. Such a value for $\xi$ is easily accommodated with the Standard Model degrees of freedom. Smaller $m_H$ or $\alpha_D$ require significantly lower $\xi$. Barring other mechanisms of entropy injection into the visible sector, $\xi=0.1(0.02)$ requires $g_{*s}^{\rm vis}(T_{\rm dec}) \sim 10^4(10^6)$.

Even with our conservative assumptions on $M_{\rm min}$, we see that the halo of dark matter that encompasses a galaxy such as the MW could contain collapsed structures with masses ranging from the equivalent of supermassive stars ($10^{2-3}M_\odot$) up to that of dwarf galaxies.

In the Standard Model, a cloud of gas with the mass of a galaxy will cool and fragment into smaller masses. The exact process is greatly complicated by the existence of the rich structure of baryonic bound states, but is ultimately stopped by nuclear fusion. Without it, the infalling gas is expected to form one or more black holes \cite{DAmico:2017lqj}. If the dark sector physics allows the existence of something akin to fusion, one could speculate about the existence of pressure-supported dark matter conglomerations, with mass scales set by additional physics which was not necessary to specify in our simple model. A new source of energy injection into the dark sector could modify the density profile of small halos while leaving larger ones unaffected. This might have interesting consequences for the structure of dwarf galaxies \cite{GarrisonKimmel:2013aq,Weinberg:2013aya,Pontzen:2011ty,Zolotov:2012xd,Governato:2012fa,Teyssier:2012ie,Brooks:2012vi,Brooks:2012ah,Arraki:2012bu,Amorisco:2013uwa}. Taking the benchmark point, with $m_H=1200$ GeV, $\alpha_D=0.1$, $m_L=10^{-3}$ GeV, and $M=10^8M_\odot$, we find $t_c \approx 0.01$ Gyr, $t_f \approx 0.09$ Gyr, and Hubble time $t_H \approx 0.7$ Gyr at $z=10$. Thus, without a significant source of energy, we should expect these objects to collapse in isolation sufficiently quickly and to become sufficiently compact to avoid significant tidal stripping as they are accreted onto larger, uncollapsed dark matter halos. They may still be susceptible to stripping if they collapse into individual disks, however dark-magnetic fields are likely to be generated in this model, providing an avenue for angular momentum loss \cite{Ackerman:mha}.

Due to tidal stripping, most subhalos falling into a MW-sized galaxy do not survive to the present day. $N$-body simulations of MW-mass galaxies finds $10-100$ ($10^5$) substructures with masses around $10^8$ ($10^4$) $M_\odot$ present in the halo at $z=0$. In all, less than 2\% of the dark matter in the MW would be in substructure \cite{Madau:2008fr}. However, only ${\sim}5\%$ of the MW was smoothly accreted from dark matter not contained in a halo \cite{2010MNRAS.401.1796A}. Instead, much of the dark matter in the MW was originally accreted as part of a smaller virialized halo; if those halos can cool, then we should expect those objects to have collapsed and retained most if not all of their dark matter. For example, if dark matter halos in the mass range $10^{7-8}M_\odot$ can cool and collapse (as realized in the parameter point $m_H = 1.2$~TeV, $m_L = 1$~MeV, and $\alpha_D = 0.1$), then fully 10\% of the MW's dark matter would be in ${\sim} 2000$ objects in this mass range \cite{Rashkov:2011cq}. Though simulations are resolution-limited for smaller dark matter halos, each decade of mass for smaller subhalos might be expected to contribute a similar fraction of the MW's mass. In a dwarf galaxy, the percentage of the total halo dark matter residing in collapsed objects could be much higher.

Dark matter forming compact objects should remind one of the massive compact halo object (MACHO) paradigm, and several of the same dynamical limits apply. In particular, tidal disruption of star clusters \cite{Brandt:2016aco}, dynamical friction in the halo \cite{Carr:1997cn}, and millilensing of quasars \cite{Wilkinson:2001vv} can exclude the possibility of 10\% of the dark matter being in primordial black holes with a monochromatic mass spectrum in the range of $10^{3-9}M_\odot$. However, the limits on collapsed objects with a broad spectrum of masses, as would be expected in our model, are generally weaker but can vary in strength depending on the halo masses \cite{Carr:2016drx,Carr:2017jsz,Poulin:2017bwe}. Furthermore, it is unclear whether these objects will collapse down to a single object, or continue to fragment down, with  smaller fragments possibly stripped from the originating subhalo (perhaps forming a ``dark disk'' \cite{Fan:2013tia,Fan:2013yva}).

In addition to constraints from disruption of galactic systems, searches for substructure using gravitational lensing will eventually probe the mass range of $10^8M_\odot$ \cite{Cyr-Racine:2015jwa}, while Lyman-$\alpha$ constraints set the best current bound at a similar mass \cite{Viel:2005qj,Seljak:2006bg} (though care must be taken when extrapolating these limits to models far removed from collisionless CDM).  Being more compact than the originating halo, a collapsed dark matter structure moving through a stellar stream in the Galaxy would presumably have a very different signature than expected from CDM \cite{Ibata:2000pu,Johnston:2001wh,Erkal:2015kqa,Bovy:2016chl}. The presence of additional collapsed structure might also affect the evolution of baryonic structures, which might be desirable in some cases (e.g.~globular clusters \cite{2014CQGra..31x4006K}). The precise details of when collapse begins and the rate of collapse can impact star formation, which may then be subject to searches for new star populations and reionization constraints \cite{Adam:2016hgk}. In all cases, observable effects could be very dependent on the exact structure of the final state, not just on the mass of the originating cooling halo. More work is needed to explore all the possibilities. 

Many particle physics models of dark matter imply that dark matter in a galactic halo is a relatively smooth distribution of (nearly) non-interacting particles. Such solutions have been very successful at matching what little is known of the dark sector. However, our own sector of physics is far from simple, and leads to a complicated evolution from the initial primordial density perturbations to galaxies, stars, and planets. While experimentally the dark sector cannot have identical parameters to our own, it is notable that no current set of measurements can rule out the possibility that dark matter also underwent an equally complex evolution, leading to collapsed structures made of dark matter at scales somewhat smaller than the MW.

In this Letter, we demonstrated that collapse can occur in an extremely simple dark matter model, consisting of two oppositely charged states and a $U(1)$ gauge force. While this simple scenario is the natural starting point, it opens some very interesting avenues for further research. These questions include: 1) Is such behavior generic for dark matter models with interactions? 2) What are the possible final states of the collapse? 3) Experimentally, what is the maximum size of dark matter halos that can collapse? and 4) How can we detect or exclude these collapsed objects? Though we have used a very familiar physical model in our example,  one would suspect that similar phenomenology could be realized using a very alien set of dark physics.

\begin{acknowledgments}
{\bf Acknowledgements:} We thank P.~Agrawal, F.Y.~Cyr-Racine, and A.~Peters for useful comments and discussion.
\end{acknowledgments}

\bibliographystyle{apsrev4-1}
\bibliography{collapse}

\begin{thebibliography}{71}%
\makeatletter
\providecommand \@ifxundefined [1]{%
 \@ifx{#1\undefined}
}%
\providecommand \@ifnum [1]{%
 \ifnum #1\expandafter \@firstoftwo
 \else \expandafter \@secondoftwo
 \fi
}%
\providecommand \@ifx [1]{%
 \ifx #1\expandafter \@firstoftwo
 \else \expandafter \@secondoftwo
 \fi
}%
\providecommand \natexlab [1]{#1}%
\providecommand \enquote  [1]{``#1''}%
\providecommand \bibnamefont  [1]{#1}%
\providecommand \bibfnamefont [1]{#1}%
\providecommand \citenamefont [1]{#1}%
\providecommand \href@noop [0]{\@secondoftwo}%
\providecommand \href [0]{\begingroup \@sanitize@url \@href}%
\providecommand \@href[1]{\@@startlink{#1}\@@href}%
\providecommand \@@href[1]{\endgroup#1\@@endlink}%
\providecommand \@sanitize@url [0]{\catcode `\\12\catcode `\$12\catcode
  `\&12\catcode `\#12\catcode `\^12\catcode `\_12\catcode `\%12\relax}%
\providecommand \@@startlink[1]{}%
\providecommand \@@endlink[0]{}%
\providecommand \url  [0]{\begingroup\@sanitize@url \@url }%
\providecommand \@url [1]{\endgroup\@href {#1}{\urlprefix }}%
\providecommand \urlprefix  [0]{URL }%
\providecommand \Eprint [0]{\href }%
\providecommand \doibase [0]{http://dx.doi.org/}%
\providecommand \selectlanguage [0]{\@gobble}%
\providecommand \bibinfo  [0]{\@secondoftwo}%
\providecommand \bibfield  [0]{\@secondoftwo}%
\providecommand \translation [1]{[#1]}%
\providecommand \BibitemOpen [0]{}%
\providecommand \bibitemStop [0]{}%
\providecommand \bibitemNoStop [0]{.\EOS\space}%
\providecommand \EOS [0]{\spacefactor3000\relax}%
\providecommand \BibitemShut  [1]{\csname bibitem#1\endcsname}%
\let\auto@bib@innerbib\@empty
%</preamble>
\bibitem [{\citenamefont {Ade}\ \emph {et~al.}(2016)\citenamefont {Ade} \emph
  {et~al.}}]{Ade:2015xua}%
  \BibitemOpen
  \bibfield  {author} {\bibinfo {author} {\bibfnamefont {P.~A.~R.}\
  \bibnamefont {Ade}} \emph {et~al.} (\bibinfo {collaboration} {Planck}),\
  }\href {\doibase 10.1051/0004-6361/201525830} {\bibfield  {journal} {\bibinfo
   {journal} {Astron. Astrophys.}\ }\textbf {\bibinfo {volume} {594}},\
  \bibinfo {pages} {A13} (\bibinfo {year} {2016})},\ \Eprint
  {http://arxiv.org/abs/1502.01589} {arXiv:1502.01589 [astro-ph.CO]}
  \BibitemShut {NoStop}%
%%CITATION = ARXIV:1502.01589;%%
\bibitem [{\citenamefont {Thoul}\ and\ \citenamefont
  {Weinberg}(1995)}]{Thoul:1994ir}%
  \BibitemOpen
  \bibfield  {author} {\bibinfo {author} {\bibfnamefont {A.~A.}\ \bibnamefont
  {Thoul}}\ and\ \bibinfo {author} {\bibfnamefont {D.~H.}\ \bibnamefont
  {Weinberg}},\ }\href {\doibase 10.1086/175455} {\bibfield  {journal}
  {\bibinfo  {journal} {Astrophys. J.}\ }\textbf {\bibinfo {volume} {442}},\
  \bibinfo {pages} {480} (\bibinfo {year} {1995})},\ \Eprint
  {http://arxiv.org/abs/astro-ph/9410009} {arXiv:astro-ph/9410009 [astro-ph]}
  \BibitemShut {NoStop}%
%%CITATION = ASTRO-PH/9410009;%%
\bibitem [{\citenamefont {{Zel'dovich}}\ and\ \citenamefont
  {{Novikov}}(1967)}]{russiansPBH}%
  \BibitemOpen
  \bibfield  {author} {\bibinfo {author} {\bibfnamefont {Y.~B.}\ \bibnamefont
  {{Zel'dovich}}}\ and\ \bibinfo {author} {\bibfnamefont {I.~D.}\ \bibnamefont
  {{Novikov}}},\ }\href@noop {} {\bibfield  {journal} {\bibinfo  {journal}
  {Sov. Astron.}\ }\textbf {\bibinfo {volume} {10}},\ \bibinfo {pages} {602}
  (\bibinfo {year} {1967})}\BibitemShut {NoStop}%
\bibitem [{\citenamefont {Hawking}(1974)}]{Hawking:1974rv}%
  \BibitemOpen
  \bibfield  {author} {\bibinfo {author} {\bibfnamefont {S.~W.}\ \bibnamefont
  {Hawking}},\ }\href {\doibase 10.1038/248030a0} {\bibfield  {journal}
  {\bibinfo  {journal} {Nature}\ }\textbf {\bibinfo {volume} {248}},\ \bibinfo
  {pages} {30} (\bibinfo {year} {1974})}\BibitemShut {NoStop}%
%%CITATION = NATUA,248,30;%%
\bibitem [{\citenamefont {Ivanov}\ \emph {et~al.}(1994)\citenamefont {Ivanov},
  \citenamefont {Naselsky},\ and\ \citenamefont {Novikov}}]{Ivanov:1994pa}%
  \BibitemOpen
  \bibfield  {author} {\bibinfo {author} {\bibfnamefont {P.}~\bibnamefont
  {Ivanov}}, \bibinfo {author} {\bibfnamefont {P.}~\bibnamefont {Naselsky}}, \
  and\ \bibinfo {author} {\bibfnamefont {I.}~\bibnamefont {Novikov}},\ }\href
  {\doibase 10.1103/PhysRevD.50.7173} {\bibfield  {journal} {\bibinfo
  {journal} {Phys. Rev.}\ }\textbf {\bibinfo {volume} {D50}},\ \bibinfo {pages}
  {7173} (\bibinfo {year} {1994})}\BibitemShut {NoStop}%
%%CITATION = PHRVA,D50,7173;%%
\bibitem [{\citenamefont {Tkachev}(1991)}]{Tkachev:1991ka}%
  \BibitemOpen
  \bibfield  {author} {\bibinfo {author} {\bibfnamefont {I.~I.}\ \bibnamefont
  {Tkachev}},\ }\href {\doibase 10.1016/0370-2693(91)90330-S} {\bibfield
  {journal} {\bibinfo  {journal} {Phys. Lett.}\ }\textbf {\bibinfo {volume}
  {B261}},\ \bibinfo {pages} {289} (\bibinfo {year} {1991})}\BibitemShut
  {NoStop}%
%%CITATION = PHLTA,B261,289;%%
\bibitem [{\citenamefont {Fan}\ \emph {et~al.}(2013{\natexlab{a}})\citenamefont
  {Fan}, \citenamefont {Katz}, \citenamefont {Randall},\ and\ \citenamefont
  {Reece}}]{Fan:2013tia}%
  \BibitemOpen
  \bibfield  {author} {\bibinfo {author} {\bibfnamefont {J.}~\bibnamefont
  {Fan}}, \bibinfo {author} {\bibfnamefont {A.}~\bibnamefont {Katz}}, \bibinfo
  {author} {\bibfnamefont {L.}~\bibnamefont {Randall}}, \ and\ \bibinfo
  {author} {\bibfnamefont {M.}~\bibnamefont {Reece}},\ }\href {\doibase
  10.1103/PhysRevLett.110.211302} {\bibfield  {journal} {\bibinfo  {journal}
  {Phys. Rev. Lett.}\ }\textbf {\bibinfo {volume} {110}},\ \bibinfo {pages}
  {211302} (\bibinfo {year} {2013}{\natexlab{a}})},\ \Eprint
  {http://arxiv.org/abs/1303.3271} {arXiv:1303.3271 [hep-ph]} \BibitemShut
  {NoStop}%
%%CITATION = ARXIV:1303.3271;%%
\bibitem [{\citenamefont {Fan}\ \emph {et~al.}(2013{\natexlab{b}})\citenamefont
  {Fan}, \citenamefont {Katz}, \citenamefont {Randall},\ and\ \citenamefont
  {Reece}}]{Fan:2013yva}%
  \BibitemOpen
  \bibfield  {author} {\bibinfo {author} {\bibfnamefont {J.}~\bibnamefont
  {Fan}}, \bibinfo {author} {\bibfnamefont {A.}~\bibnamefont {Katz}}, \bibinfo
  {author} {\bibfnamefont {L.}~\bibnamefont {Randall}}, \ and\ \bibinfo
  {author} {\bibfnamefont {M.}~\bibnamefont {Reece}},\ }\href {\doibase
  10.1016/j.dark.2013.07.001} {\bibfield  {journal} {\bibinfo  {journal} {Phys.
  Dark Univ.}\ }\textbf {\bibinfo {volume} {2}},\ \bibinfo {pages} {139}
  (\bibinfo {year} {2013}{\natexlab{b}})},\ \Eprint
  {http://arxiv.org/abs/1303.1521} {arXiv:1303.1521 [astro-ph.CO]} \BibitemShut
  {NoStop}%
%%CITATION = ARXIV:1303.1521;%%
\bibitem [{\citenamefont {Agrawal}\ \emph
  {et~al.}(2017{\natexlab{a}})\citenamefont {Agrawal}, \citenamefont
  {Cyr-Racine}, \citenamefont {Randall},\ and\ \citenamefont
  {Scholtz}}]{Agrawal:2017rvu}%
  \BibitemOpen
  \bibfield  {author} {\bibinfo {author} {\bibfnamefont {P.}~\bibnamefont
  {Agrawal}}, \bibinfo {author} {\bibfnamefont {F.-Y.}\ \bibnamefont
  {Cyr-Racine}}, \bibinfo {author} {\bibfnamefont {L.}~\bibnamefont {Randall}},
  \ and\ \bibinfo {author} {\bibfnamefont {J.}~\bibnamefont {Scholtz}},\
  }\href@noop {} {\  (\bibinfo {year} {2017}{\natexlab{a}})},\ \Eprint
  {http://arxiv.org/abs/1702.05482} {arXiv:1702.05482 [astro-ph.CO]}
  \BibitemShut {NoStop}%
%%CITATION = ARXIV:1702.05482;%%
\bibitem [{\citenamefont {Fischler}\ \emph {et~al.}(2015)\citenamefont
  {Fischler}, \citenamefont {Lorshbough},\ and\ \citenamefont
  {Tangarife}}]{Fischler:2014jda}%
  \BibitemOpen
  \bibfield  {author} {\bibinfo {author} {\bibfnamefont {W.}~\bibnamefont
  {Fischler}}, \bibinfo {author} {\bibfnamefont {D.}~\bibnamefont
  {Lorshbough}}, \ and\ \bibinfo {author} {\bibfnamefont {W.}~\bibnamefont
  {Tangarife}},\ }\href {\doibase 10.1103/PhysRevD.91.025010} {\bibfield
  {journal} {\bibinfo  {journal} {Phys. Rev.}\ }\textbf {\bibinfo {volume}
  {D91}},\ \bibinfo {pages} {025010} (\bibinfo {year} {2015})},\ \Eprint
  {http://arxiv.org/abs/1405.7708} {arXiv:1405.7708 [hep-ph]} \BibitemShut
  {NoStop}%
%%CITATION = ARXIV:1405.7708;%%
\bibitem [{\citenamefont {Mohapatra}\ and\ \citenamefont
  {Teplitz}(1997)}]{Mohapatra:1996yy}%
  \BibitemOpen
  \bibfield  {author} {\bibinfo {author} {\bibfnamefont {R.~N.}\ \bibnamefont
  {Mohapatra}}\ and\ \bibinfo {author} {\bibfnamefont {V.~L.}\ \bibnamefont
  {Teplitz}},\ }\href {\doibase 10.1086/303762} {\bibfield  {journal} {\bibinfo
   {journal} {Astrophys. J.}\ }\textbf {\bibinfo {volume} {478}},\ \bibinfo
  {pages} {29} (\bibinfo {year} {1997})},\ \Eprint
  {http://arxiv.org/abs/astro-ph/9603049} {arXiv:astro-ph/9603049 [astro-ph]}
  \BibitemShut {NoStop}%
%%CITATION = ASTRO-PH/9603049;%%
\bibitem [{\citenamefont {Tulin}\ and\ \citenamefont
  {Yu}(2017)}]{Tulin:2017ara}%
  \BibitemOpen
  \bibfield  {author} {\bibinfo {author} {\bibfnamefont {S.}~\bibnamefont
  {Tulin}}\ and\ \bibinfo {author} {\bibfnamefont {H.-B.}\ \bibnamefont {Yu}},\
  }\href@noop {} {\  (\bibinfo {year} {2017})},\ \Eprint
  {http://arxiv.org/abs/1705.02358} {arXiv:1705.02358 [hep-ph]} \BibitemShut
  {NoStop}%
%%CITATION = ARXIV:1705.02358;%%
\bibitem [{\citenamefont {Viel}\ \emph {et~al.}(2005)\citenamefont {Viel},
  \citenamefont {Lesgourgues}, \citenamefont {Haehnelt}, \citenamefont
  {Matarrese},\ and\ \citenamefont {Riotto}}]{Viel:2005qj}%
  \BibitemOpen
  \bibfield  {author} {\bibinfo {author} {\bibfnamefont {M.}~\bibnamefont
  {Viel}}, \bibinfo {author} {\bibfnamefont {J.}~\bibnamefont {Lesgourgues}},
  \bibinfo {author} {\bibfnamefont {M.~G.}\ \bibnamefont {Haehnelt}}, \bibinfo
  {author} {\bibfnamefont {S.}~\bibnamefont {Matarrese}}, \ and\ \bibinfo
  {author} {\bibfnamefont {A.}~\bibnamefont {Riotto}},\ }\href {\doibase
  10.1103/PhysRevD.71.063534} {\bibfield  {journal} {\bibinfo  {journal} {Phys.
  Rev.}\ }\textbf {\bibinfo {volume} {D71}},\ \bibinfo {pages} {063534}
  (\bibinfo {year} {2005})},\ \Eprint {http://arxiv.org/abs/astro-ph/0501562}
  {arXiv:astro-ph/0501562 [astro-ph]} \BibitemShut {NoStop}%
%%CITATION = ASTRO-PH/0501562;%%
\bibitem [{\citenamefont {Seljak}\ \emph {et~al.}(2006)\citenamefont {Seljak},
  \citenamefont {Slosar},\ and\ \citenamefont {McDonald}}]{Seljak:2006bg}%
  \BibitemOpen
  \bibfield  {author} {\bibinfo {author} {\bibfnamefont {U.}~\bibnamefont
  {Seljak}}, \bibinfo {author} {\bibfnamefont {A.}~\bibnamefont {Slosar}}, \
  and\ \bibinfo {author} {\bibfnamefont {P.}~\bibnamefont {McDonald}},\ }\href
  {\doibase 10.1088/1475-7516/2006/10/014} {\bibfield  {journal} {\bibinfo
  {journal} {JCAP}\ }\textbf {\bibinfo {volume} {0610}},\ \bibinfo {pages}
  {014} (\bibinfo {year} {2006})},\ \Eprint
  {http://arxiv.org/abs/astro-ph/0604335} {arXiv:astro-ph/0604335 [astro-ph]}
  \BibitemShut {NoStop}%
%%CITATION = ASTRO-PH/0604335;%%
\bibitem [{\citenamefont {Silk}(1977)}]{Silk:1977wz}%
  \BibitemOpen
  \bibfield  {author} {\bibinfo {author} {\bibfnamefont {J.}~\bibnamefont
  {Silk}},\ }\href {\doibase 10.1086/154972} {\bibfield  {journal} {\bibinfo
  {journal} {Astrophys. J.}\ }\textbf {\bibinfo {volume} {211}},\ \bibinfo
  {pages} {638} (\bibinfo {year} {1977})}\BibitemShut {NoStop}%
%%CITATION = ASJOA,211,638;%%
\bibitem [{\citenamefont {White}\ and\ \citenamefont
  {Rees}(1978)}]{White:1977jf}%
  \BibitemOpen
  \bibfield  {author} {\bibinfo {author} {\bibfnamefont {S.~D.~M.}\
  \bibnamefont {White}}\ and\ \bibinfo {author} {\bibfnamefont {M.~J.}\
  \bibnamefont {Rees}},\ }\href@noop {} {\bibfield  {journal} {\bibinfo
  {journal} {Mon. Not. Roy. Astron. Soc.}\ }\textbf {\bibinfo {volume} {183}},\
  \bibinfo {pages} {341} (\bibinfo {year} {1978})}\BibitemShut {NoStop}%
%%CITATION = MNRAA,183,341;%%
\bibitem [{\citenamefont {Silk}\ \emph {et~al.}(2014)\citenamefont {Silk},
  \citenamefont {Di~Cintio},\ and\ \citenamefont {Dvorkin}}]{Silk:2013xca}%
  \BibitemOpen
  \bibfield  {author} {\bibinfo {author} {\bibfnamefont {J.}~\bibnamefont
  {Silk}}, \bibinfo {author} {\bibfnamefont {A.}~\bibnamefont {Di~Cintio}}, \
  and\ \bibinfo {author} {\bibfnamefont {I.}~\bibnamefont {Dvorkin}},\
  }\bibfield  {booktitle} {\emph {\bibinfo {booktitle} {{Proceedings,
  International School of Physics 'Enrico Fermi': New Horizons for
  Observational Cosmology: Rome, Italy, June 30-July 6, 2013}}},\ }\href
  {\doibase 10.3254/978-1-61499-476-3-137,
  10.1093/acprof:oso/9780198728856.003.0009} {\bibfield  {journal} {\bibinfo
  {journal} {Proc. Int. Sch. Phys. Fermi}\ }\textbf {\bibinfo {volume} {186}},\
  \bibinfo {pages} {137} (\bibinfo {year} {2014})},\ \bibinfo {note}
  {[,407(2015)]},\ \Eprint {http://arxiv.org/abs/1312.0107} {arXiv:1312.0107
  [astro-ph.CO]} \BibitemShut {NoStop}%
%%CITATION = ARXIV:1312.0107;%%
\bibitem [{\citenamefont {Ackerman}\ \emph {et~al.}(2009)\citenamefont
  {Ackerman}, \citenamefont {Buckley}, \citenamefont {Carroll},\ and\
  \citenamefont {Kamionkowski}}]{Ackerman:mha}%
  \BibitemOpen
  \bibfield  {author} {\bibinfo {author} {\bibfnamefont {L.}~\bibnamefont
  {Ackerman}}, \bibinfo {author} {\bibfnamefont {M.~R.}\ \bibnamefont
  {Buckley}}, \bibinfo {author} {\bibfnamefont {S.~M.}\ \bibnamefont
  {Carroll}}, \ and\ \bibinfo {author} {\bibfnamefont {M.}~\bibnamefont
  {Kamionkowski}},\ }\bibfield  {booktitle} {\emph {\bibinfo {booktitle}
  {{Proceedings, 7th International Heidelberg Conference on Dark Matter in
  Astro and Particle Physics (DARK 2009): Christchurch, New Zealand, January
  18-24, 2009}}},\ }\href {\doibase 10.1103/PhysRevD.79.023519,
  10.1142/9789814293792_0021} {\bibfield  {journal} {\bibinfo  {journal} {Phys.
  Rev.}\ }\textbf {\bibinfo {volume} {D79}},\ \bibinfo {pages} {023519}
  (\bibinfo {year} {2009})},\ \bibinfo {note} {[,277(2008)]},\ \Eprint
  {http://arxiv.org/abs/0810.5126} {arXiv:0810.5126 [hep-ph]} \BibitemShut
  {NoStop}%
%%CITATION = ARXIV:0810.5126;%%
\bibitem [{\citenamefont {Zurek}(2014)}]{Zurek:2013wia}%
  \BibitemOpen
  \bibfield  {author} {\bibinfo {author} {\bibfnamefont {K.~M.}\ \bibnamefont
  {Zurek}},\ }\href {\doibase 10.1016/j.physrep.2013.12.001} {\bibfield
  {journal} {\bibinfo  {journal} {Phys. Rept.}\ }\textbf {\bibinfo {volume}
  {537}},\ \bibinfo {pages} {91} (\bibinfo {year} {2014})},\ \Eprint
  {http://arxiv.org/abs/1308.0338} {arXiv:1308.0338 [hep-ph]} \BibitemShut
  {NoStop}%
%%CITATION = ARXIV:1308.0338;%%
\bibitem [{\citenamefont {Buckley}(2011)}]{Buckley:2011kk}%
  \BibitemOpen
  \bibfield  {author} {\bibinfo {author} {\bibfnamefont {M.~R.}\ \bibnamefont
  {Buckley}},\ }\href {\doibase 10.1103/PhysRevD.84.043510} {\bibfield
  {journal} {\bibinfo  {journal} {Phys. Rev.}\ }\textbf {\bibinfo {volume}
  {D84}},\ \bibinfo {pages} {043510} (\bibinfo {year} {2011})},\ \Eprint
  {http://arxiv.org/abs/1104.1429} {arXiv:1104.1429 [hep-ph]} \BibitemShut
  {NoStop}%
%%CITATION = ARXIV:1104.1429;%%
\bibitem [{\citenamefont {Feng}\ \emph {et~al.}(2009)\citenamefont {Feng},
  \citenamefont {Kaplinghat}, \citenamefont {Tu},\ and\ \citenamefont
  {Yu}}]{Feng:2009mn}%
  \BibitemOpen
  \bibfield  {author} {\bibinfo {author} {\bibfnamefont {J.~L.}\ \bibnamefont
  {Feng}}, \bibinfo {author} {\bibfnamefont {M.}~\bibnamefont {Kaplinghat}},
  \bibinfo {author} {\bibfnamefont {H.}~\bibnamefont {Tu}}, \ and\ \bibinfo
  {author} {\bibfnamefont {H.-B.}\ \bibnamefont {Yu}},\ }\href {\doibase
  10.1088/1475-7516/2009/07/004} {\bibfield  {journal} {\bibinfo  {journal}
  {JCAP}\ }\textbf {\bibinfo {volume} {0907}},\ \bibinfo {pages} {004}
  (\bibinfo {year} {2009})},\ \Eprint {http://arxiv.org/abs/0905.3039}
  {arXiv:0905.3039 [hep-ph]} \BibitemShut {NoStop}%
%%CITATION = ARXIV:0905.3039;%%
\bibitem [{\citenamefont {Kaplan}\ \emph {et~al.}(2010)\citenamefont {Kaplan},
  \citenamefont {Krnjaic}, \citenamefont {Rehermann},\ and\ \citenamefont
  {Wells}}]{Kaplan:2009de}%
  \BibitemOpen
  \bibfield  {author} {\bibinfo {author} {\bibfnamefont {D.~E.}\ \bibnamefont
  {Kaplan}}, \bibinfo {author} {\bibfnamefont {G.~Z.}\ \bibnamefont {Krnjaic}},
  \bibinfo {author} {\bibfnamefont {K.~R.}\ \bibnamefont {Rehermann}}, \ and\
  \bibinfo {author} {\bibfnamefont {C.~M.}\ \bibnamefont {Wells}},\ }\href
  {\doibase 10.1088/1475-7516/2010/05/021} {\bibfield  {journal} {\bibinfo
  {journal} {JCAP}\ }\textbf {\bibinfo {volume} {1005}},\ \bibinfo {pages}
  {021} (\bibinfo {year} {2010})},\ \Eprint {http://arxiv.org/abs/0909.0753}
  {arXiv:0909.0753 [hep-ph]} \BibitemShut {NoStop}%
%%CITATION = ARXIV:0909.0753;%%
\bibitem [{\citenamefont {Cyr-Racine}\ and\ \citenamefont
  {Sigurdson}(2013)}]{CyrRacine:2012fz}%
  \BibitemOpen
  \bibfield  {author} {\bibinfo {author} {\bibfnamefont {F.-Y.}\ \bibnamefont
  {Cyr-Racine}}\ and\ \bibinfo {author} {\bibfnamefont {K.}~\bibnamefont
  {Sigurdson}},\ }\href {\doibase 10.1103/PhysRevD.87.103515} {\bibfield
  {journal} {\bibinfo  {journal} {Phys. Rev.}\ }\textbf {\bibinfo {volume}
  {D87}},\ \bibinfo {pages} {103515} (\bibinfo {year} {2013})},\ \Eprint
  {http://arxiv.org/abs/1209.5752} {arXiv:1209.5752 [astro-ph.CO]} \BibitemShut
  {NoStop}%
%%CITATION = ARXIV:1209.5752;%%
\bibitem [{\citenamefont {Foot}(2014)}]{Foot:2014mia}%
  \BibitemOpen
  \bibfield  {author} {\bibinfo {author} {\bibfnamefont {R.}~\bibnamefont
  {Foot}},\ }\href {\doibase 10.1142/S0217751X14300130} {\bibfield  {journal}
  {\bibinfo  {journal} {Int. J. Mod. Phys.}\ }\textbf {\bibinfo {volume}
  {A29}},\ \bibinfo {pages} {1430013} (\bibinfo {year} {2014})},\ \Eprint
  {http://arxiv.org/abs/1401.3965} {arXiv:1401.3965 [astro-ph.CO]} \BibitemShut
  {NoStop}%
%%CITATION = ARXIV:1401.3965;%%
\bibitem [{\citenamefont {Foot}\ and\ \citenamefont
  {Vagnozzi}(2015)}]{Foot:2014uba}%
  \BibitemOpen
  \bibfield  {author} {\bibinfo {author} {\bibfnamefont {R.}~\bibnamefont
  {Foot}}\ and\ \bibinfo {author} {\bibfnamefont {S.}~\bibnamefont
  {Vagnozzi}},\ }\href {\doibase 10.1103/PhysRevD.91.023512} {\bibfield
  {journal} {\bibinfo  {journal} {Phys. Rev.}\ }\textbf {\bibinfo {volume}
  {D91}},\ \bibinfo {pages} {023512} (\bibinfo {year} {2015})},\ \Eprint
  {http://arxiv.org/abs/1409.7174} {arXiv:1409.7174 [hep-ph]} \BibitemShut
  {NoStop}%
%%CITATION = ARXIV:1409.7174;%%
\bibitem [{\citenamefont {Boddy}\ \emph {et~al.}(2016)\citenamefont {Boddy},
  \citenamefont {Kaplinghat}, \citenamefont {Kwa},\ and\ \citenamefont
  {Peter}}]{Boddy:2016bbu}%
  \BibitemOpen
  \bibfield  {author} {\bibinfo {author} {\bibfnamefont {K.~K.}\ \bibnamefont
  {Boddy}}, \bibinfo {author} {\bibfnamefont {M.}~\bibnamefont {Kaplinghat}},
  \bibinfo {author} {\bibfnamefont {A.}~\bibnamefont {Kwa}}, \ and\ \bibinfo
  {author} {\bibfnamefont {A.~H.~G.}\ \bibnamefont {Peter}},\ }\href {\doibase
  10.1103/PhysRevD.94.123017} {\bibfield  {journal} {\bibinfo  {journal} {Phys.
  Rev.}\ }\textbf {\bibinfo {volume} {D94}},\ \bibinfo {pages} {123017}
  (\bibinfo {year} {2016})},\ \Eprint {http://arxiv.org/abs/1609.03592}
  {arXiv:1609.03592 [hep-ph]} \BibitemShut {NoStop}%
%%CITATION = ARXIV:1609.03592;%%
\bibitem [{\citenamefont {Rosenberg}\ and\ \citenamefont
  {Fan}(2017)}]{Rosenberg:2017qia}%
  \BibitemOpen
  \bibfield  {author} {\bibinfo {author} {\bibfnamefont {E.}~\bibnamefont
  {Rosenberg}}\ and\ \bibinfo {author} {\bibfnamefont {J.}~\bibnamefont
  {Fan}},\ }\href@noop {} {\  (\bibinfo {year} {2017})},\ \Eprint
  {http://arxiv.org/abs/1705.10341} {arXiv:1705.10341 [astro-ph.GA]}
  \BibitemShut {NoStop}%
%%CITATION = ARXIV:1705.10341;%%
\bibitem [{\citenamefont {Cyr-Racine}\ \emph {et~al.}(2014)\citenamefont
  {Cyr-Racine}, \citenamefont {de~Putter}, \citenamefont {Raccanelli},\ and\
  \citenamefont {Sigurdson}}]{Cyr-Racine:2013fsa}%
  \BibitemOpen
  \bibfield  {author} {\bibinfo {author} {\bibfnamefont {F.-Y.}\ \bibnamefont
  {Cyr-Racine}}, \bibinfo {author} {\bibfnamefont {R.}~\bibnamefont
  {de~Putter}}, \bibinfo {author} {\bibfnamefont {A.}~\bibnamefont
  {Raccanelli}}, \ and\ \bibinfo {author} {\bibfnamefont {K.}~\bibnamefont
  {Sigurdson}},\ }\href {\doibase 10.1103/PhysRevD.89.063517} {\bibfield
  {journal} {\bibinfo  {journal} {Phys. Rev.}\ }\textbf {\bibinfo {volume}
  {D89}},\ \bibinfo {pages} {063517} (\bibinfo {year} {2014})},\ \Eprint
  {http://arxiv.org/abs/1310.3278} {arXiv:1310.3278 [astro-ph.CO]} \BibitemShut
  {NoStop}%
%%CITATION = ARXIV:1310.3278;%%
\bibitem [{\citenamefont {Agrawal}\ and\ \citenamefont
  {Randall}(2017)}]{Agrawal:2017pnb}%
  \BibitemOpen
  \bibfield  {author} {\bibinfo {author} {\bibfnamefont {P.}~\bibnamefont
  {Agrawal}}\ and\ \bibinfo {author} {\bibfnamefont {L.}~\bibnamefont
  {Randall}},\ }\href@noop {} {\  (\bibinfo {year} {2017})},\ \Eprint
  {http://arxiv.org/abs/1706.04195} {arXiv:1706.04195 [hep-ph]} \BibitemShut
  {NoStop}%
%%CITATION = ARXIV:1706.04195;%%
\bibitem [{\citenamefont {Berezhiani}\ \emph {et~al.}(1996)\citenamefont
  {Berezhiani}, \citenamefont {Dolgov},\ and\ \citenamefont
  {Mohapatra}}]{Berezhiani:1995am}%
  \BibitemOpen
  \bibfield  {author} {\bibinfo {author} {\bibfnamefont {Z.~G.}\ \bibnamefont
  {Berezhiani}}, \bibinfo {author} {\bibfnamefont {A.~D.}\ \bibnamefont
  {Dolgov}}, \ and\ \bibinfo {author} {\bibfnamefont {R.~N.}\ \bibnamefont
  {Mohapatra}},\ }\href {\doibase 10.1016/0370-2693(96)00219-5} {\bibfield
  {journal} {\bibinfo  {journal} {Phys. Lett.}\ }\textbf {\bibinfo {volume}
  {B375}},\ \bibinfo {pages} {26} (\bibinfo {year} {1996})},\ \Eprint
  {http://arxiv.org/abs/hep-ph/9511221} {arXiv:hep-ph/9511221 [hep-ph]}
  \BibitemShut {NoStop}%
%%CITATION = HEP-PH/9511221;%%
\bibitem [{\citenamefont {Hall}\ \emph {et~al.}(2010)\citenamefont {Hall},
  \citenamefont {Jedamzik}, \citenamefont {March-Russell},\ and\ \citenamefont
  {West}}]{Hall:2009bx}%
  \BibitemOpen
  \bibfield  {author} {\bibinfo {author} {\bibfnamefont {L.~J.}\ \bibnamefont
  {Hall}}, \bibinfo {author} {\bibfnamefont {K.}~\bibnamefont {Jedamzik}},
  \bibinfo {author} {\bibfnamefont {J.}~\bibnamefont {March-Russell}}, \ and\
  \bibinfo {author} {\bibfnamefont {S.~M.}\ \bibnamefont {West}},\ }\href
  {\doibase 10.1007/JHEP03(2010)080} {\bibfield  {journal} {\bibinfo  {journal}
  {JHEP}\ }\textbf {\bibinfo {volume} {03}},\ \bibinfo {pages} {080} (\bibinfo
  {year} {2010})},\ \Eprint {http://arxiv.org/abs/0911.1120} {arXiv:0911.1120
  [hep-ph]} \BibitemShut {NoStop}%
%%CITATION = ARXIV:0911.1120;%%
\bibitem [{\citenamefont {Allahverdi}\ \emph {et~al.}(2010)\citenamefont
  {Allahverdi}, \citenamefont {Brandenberger}, \citenamefont {Cyr-Racine},\
  and\ \citenamefont {Mazumdar}}]{Allahverdi:2010xz}%
  \BibitemOpen
  \bibfield  {author} {\bibinfo {author} {\bibfnamefont {R.}~\bibnamefont
  {Allahverdi}}, \bibinfo {author} {\bibfnamefont {R.}~\bibnamefont
  {Brandenberger}}, \bibinfo {author} {\bibfnamefont {F.-Y.}\ \bibnamefont
  {Cyr-Racine}}, \ and\ \bibinfo {author} {\bibfnamefont {A.}~\bibnamefont
  {Mazumdar}},\ }\href {\doibase 10.1146/annurev.nucl.012809.104511} {\bibfield
   {journal} {\bibinfo  {journal} {Ann. Rev. Nucl. Part. Sci.}\ }\textbf
  {\bibinfo {volume} {60}},\ \bibinfo {pages} {27} (\bibinfo {year} {2010})},\
  \Eprint {http://arxiv.org/abs/1001.2600} {arXiv:1001.2600 [hep-th]}
  \BibitemShut {NoStop}%
%%CITATION = ARXIV:1001.2600;%%
\bibitem [{\citenamefont {Adshead}\ \emph {et~al.}(2016)\citenamefont
  {Adshead}, \citenamefont {Cui},\ and\ \citenamefont
  {Shelton}}]{Adshead:2016xxj}%
  \BibitemOpen
  \bibfield  {author} {\bibinfo {author} {\bibfnamefont {P.}~\bibnamefont
  {Adshead}}, \bibinfo {author} {\bibfnamefont {Y.}~\bibnamefont {Cui}}, \ and\
  \bibinfo {author} {\bibfnamefont {J.}~\bibnamefont {Shelton}},\ }\href
  {\doibase 10.1007/JHEP06(2016)016} {\bibfield  {journal} {\bibinfo  {journal}
  {JHEP}\ }\textbf {\bibinfo {volume} {06}},\ \bibinfo {pages} {016} (\bibinfo
  {year} {2016})},\ \Eprint {http://arxiv.org/abs/1604.02458} {arXiv:1604.02458
  [hep-ph]} \BibitemShut {NoStop}%
%%CITATION = ARXIV:1604.02458;%%
\bibitem [{\citenamefont {Bardeen}\ \emph {et~al.}(1983)\citenamefont
  {Bardeen}, \citenamefont {Steinhardt},\ and\ \citenamefont
  {Turner}}]{Bardeen:1983qw}%
  \BibitemOpen
  \bibfield  {author} {\bibinfo {author} {\bibfnamefont {J.~M.}\ \bibnamefont
  {Bardeen}}, \bibinfo {author} {\bibfnamefont {P.~J.}\ \bibnamefont
  {Steinhardt}}, \ and\ \bibinfo {author} {\bibfnamefont {M.~S.}\ \bibnamefont
  {Turner}},\ }\href {\doibase 10.1103/PhysRevD.28.679} {\bibfield  {journal}
  {\bibinfo  {journal} {Phys. Rev.}\ }\textbf {\bibinfo {volume} {D28}},\
  \bibinfo {pages} {679} (\bibinfo {year} {1983})}\BibitemShut {NoStop}%
%%CITATION = PHRVA,D28,679;%%
\bibitem [{\citenamefont {Blumenthal}\ \emph {et~al.}(1984)\citenamefont
  {Blumenthal}, \citenamefont {Faber}, \citenamefont {Primack},\ and\
  \citenamefont {Rees}}]{Blumenthal:1984bp}%
  \BibitemOpen
  \bibfield  {author} {\bibinfo {author} {\bibfnamefont {G.~R.}\ \bibnamefont
  {Blumenthal}}, \bibinfo {author} {\bibfnamefont {S.~M.}\ \bibnamefont
  {Faber}}, \bibinfo {author} {\bibfnamefont {J.~R.}\ \bibnamefont {Primack}},
  \ and\ \bibinfo {author} {\bibfnamefont {M.~J.}\ \bibnamefont {Rees}},\
  }\href {\doibase 10.1038/311517a0} {\bibfield  {journal} {\bibinfo  {journal}
  {Nature}\ }\textbf {\bibinfo {volume} {311}},\ \bibinfo {pages} {517}
  (\bibinfo {year} {1984})}\BibitemShut {NoStop}%
%%CITATION = NATUA,311,517;%%
\bibitem [{\citenamefont {Springel}\ \emph {et~al.}(2005)\citenamefont
  {Springel} \emph {et~al.}}]{Springel:2005nw}%
  \BibitemOpen
  \bibfield  {author} {\bibinfo {author} {\bibfnamefont {V.}~\bibnamefont
  {Springel}} \emph {et~al.},\ }\href {\doibase 10.1038/nature03597} {\bibfield
   {journal} {\bibinfo  {journal} {Nature}\ }\textbf {\bibinfo {volume}
  {435}},\ \bibinfo {pages} {629} (\bibinfo {year} {2005})},\ \Eprint
  {http://arxiv.org/abs/astro-ph/0504097} {arXiv:astro-ph/0504097 [astro-ph]}
  \BibitemShut {NoStop}%
%%CITATION = ASTRO-PH/0504097;%%
\bibitem [{\citenamefont {Behroozi}\ and\ \citenamefont
  {Silk}(2015)}]{Behroozi:2014tna}%
  \BibitemOpen
  \bibfield  {author} {\bibinfo {author} {\bibfnamefont {P.~S.}\ \bibnamefont
  {Behroozi}}\ and\ \bibinfo {author} {\bibfnamefont {J.}~\bibnamefont
  {Silk}},\ }\href {\doibase 10.1088/0004-637X/799/1/32} {\bibfield  {journal}
  {\bibinfo  {journal} {Astrophys. J.}\ }\textbf {\bibinfo {volume} {799}},\
  \bibinfo {pages} {32} (\bibinfo {year} {2015})},\ \Eprint
  {http://arxiv.org/abs/1404.5299} {arXiv:1404.5299 [astro-ph.GA]} \BibitemShut
  {NoStop}%
%%CITATION = ARXIV:1404.5299;%%
\bibitem [{\citenamefont {Feng}\ \emph {et~al.}(2008)\citenamefont {Feng},
  \citenamefont {Tu},\ and\ \citenamefont {Yu}}]{Feng:2008mu}%
  \BibitemOpen
  \bibfield  {author} {\bibinfo {author} {\bibfnamefont {J.~L.}\ \bibnamefont
  {Feng}}, \bibinfo {author} {\bibfnamefont {H.}~\bibnamefont {Tu}}, \ and\
  \bibinfo {author} {\bibfnamefont {H.-B.}\ \bibnamefont {Yu}},\ }\href
  {\doibase 10.1088/1475-7516/2008/10/043} {\bibfield  {journal} {\bibinfo
  {journal} {JCAP}\ }\textbf {\bibinfo {volume} {0810}},\ \bibinfo {pages}
  {043} (\bibinfo {year} {2008})},\ \Eprint {http://arxiv.org/abs/0808.2318}
  {arXiv:0808.2318 [hep-ph]} \BibitemShut {NoStop}%
%%CITATION = ARXIV:0808.2318;%%
\bibitem [{\citenamefont {{Mo}}\ \emph {et~al.}(2010)\citenamefont {{Mo}},
  \citenamefont {{van den Bosch}},\ and\ \citenamefont {{White}}}]{Mo:2010gfe}%
  \BibitemOpen
  \bibfield  {author} {\bibinfo {author} {\bibfnamefont {H.}~\bibnamefont
  {{Mo}}}, \bibinfo {author} {\bibfnamefont {F.~C.}\ \bibnamefont {{van den
  Bosch}}}, \ and\ \bibinfo {author} {\bibfnamefont {S.}~\bibnamefont
  {{White}}},\ }\href@noop {} {\emph {\bibinfo {title} {Galaxy Formation and
  Evolution, by Houjun Mo , Frank van den Bosch , Simon White, Cambridge, UK:
  Cambridge University Press, 2010}}}\ (\bibinfo {year} {2010})\BibitemShut
  {NoStop}%
\bibitem [{\citenamefont {Kim}\ and\ \citenamefont
  {Rudd}(1994)}]{PhysRevA.50.3954}%
  \BibitemOpen
  \bibfield  {author} {\bibinfo {author} {\bibfnamefont {Y.-K.}\ \bibnamefont
  {Kim}}\ and\ \bibinfo {author} {\bibfnamefont {M.~E.}\ \bibnamefont {Rudd}},\
  }\href {\doibase 10.1103/PhysRevA.50.3954} {\bibfield  {journal} {\bibinfo
  {journal} {Phys. Rev. A}\ }\textbf {\bibinfo {volume} {50}},\ \bibinfo
  {pages} {3954} (\bibinfo {year} {1994})}\BibitemShut {NoStop}%
\bibitem [{\citenamefont {{Karzas}}\ and\ \citenamefont
  {{Latter}}(1961)}]{Karzas:1961ApJS}%
  \BibitemOpen
  \bibfield  {author} {\bibinfo {author} {\bibfnamefont {W.~J.}\ \bibnamefont
  {{Karzas}}}\ and\ \bibinfo {author} {\bibfnamefont {R.}~\bibnamefont
  {{Latter}}},\ }\href {\doibase 10.1086/190063} {\bibfield  {journal}
  {\bibinfo  {journal} {ApJS}\ }\textbf {\bibinfo {volume} {6}},\ \bibinfo
  {pages} {167} (\bibinfo {year} {1961})}\BibitemShut {NoStop}%
\bibitem [{\citenamefont {Agrawal}\ \emph
  {et~al.}(2017{\natexlab{b}})\citenamefont {Agrawal}, \citenamefont
  {Cyr-Racine}, \citenamefont {Randall},\ and\ \citenamefont
  {Scholtz}}]{Agrawal:2016quu}%
  \BibitemOpen
  \bibfield  {author} {\bibinfo {author} {\bibfnamefont {P.}~\bibnamefont
  {Agrawal}}, \bibinfo {author} {\bibfnamefont {F.-Y.}\ \bibnamefont
  {Cyr-Racine}}, \bibinfo {author} {\bibfnamefont {L.}~\bibnamefont {Randall}},
  \ and\ \bibinfo {author} {\bibfnamefont {J.}~\bibnamefont {Scholtz}},\ }\href
  {\doibase 10.1088/1475-7516/2017/05/022} {\bibfield  {journal} {\bibinfo
  {journal} {JCAP}\ }\textbf {\bibinfo {volume} {1705}},\ \bibinfo {pages}
  {022} (\bibinfo {year} {2017}{\natexlab{b}})},\ \Eprint
  {http://arxiv.org/abs/1610.04611} {arXiv:1610.04611 [hep-ph]} \BibitemShut
  {NoStop}%
%%CITATION = ARXIV:1610.04611;%%
\bibitem [{\citenamefont {{Silk}}(1968)}]{1968ApJ...151..459S}%
  \BibitemOpen
  \bibfield  {author} {\bibinfo {author} {\bibfnamefont {J.}~\bibnamefont
  {{Silk}}},\ }\href {\doibase 10.1086/149449} {\bibfield  {journal} {\bibinfo
  {journal} {Astrophys. J.}\ }\textbf {\bibinfo {volume} {151}},\ \bibinfo
  {pages} {459} (\bibinfo {year} {1968})}\BibitemShut {NoStop}%
\bibitem [{\citenamefont {Buckley}\ \emph {et~al.}(2014)\citenamefont
  {Buckley}, \citenamefont {Zavala}, \citenamefont {Cyr-Racine}, \citenamefont
  {Sigurdson},\ and\ \citenamefont {Vogelsberger}}]{Buckley:2014hja}%
  \BibitemOpen
  \bibfield  {author} {\bibinfo {author} {\bibfnamefont {M.~R.}\ \bibnamefont
  {Buckley}}, \bibinfo {author} {\bibfnamefont {J.}~\bibnamefont {Zavala}},
  \bibinfo {author} {\bibfnamefont {F.-Y.}\ \bibnamefont {Cyr-Racine}},
  \bibinfo {author} {\bibfnamefont {K.}~\bibnamefont {Sigurdson}}, \ and\
  \bibinfo {author} {\bibfnamefont {M.}~\bibnamefont {Vogelsberger}},\ }\href
  {\doibase 10.1103/PhysRevD.90.043524} {\bibfield  {journal} {\bibinfo
  {journal} {Phys. Rev.}\ }\textbf {\bibinfo {volume} {D90}},\ \bibinfo {pages}
  {043524} (\bibinfo {year} {2014})},\ \Eprint {http://arxiv.org/abs/1405.2075}
  {arXiv:1405.2075 [astro-ph.CO]} \BibitemShut {NoStop}%
%%CITATION = ARXIV:1405.2075;%%
\bibitem [{\citenamefont {D'Amico}\ \emph {et~al.}(2017)\citenamefont
  {D'Amico}, \citenamefont {Panci}, \citenamefont {Lupi}, \citenamefont
  {Bovino},\ and\ \citenamefont {Silk}}]{DAmico:2017lqj}%
  \BibitemOpen
  \bibfield  {author} {\bibinfo {author} {\bibfnamefont {G.}~\bibnamefont
  {D'Amico}}, \bibinfo {author} {\bibfnamefont {P.}~\bibnamefont {Panci}},
  \bibinfo {author} {\bibfnamefont {A.}~\bibnamefont {Lupi}}, \bibinfo {author}
  {\bibfnamefont {S.}~\bibnamefont {Bovino}}, \ and\ \bibinfo {author}
  {\bibfnamefont {J.}~\bibnamefont {Silk}},\ }\href@noop {} {\  (\bibinfo
  {year} {2017})},\ \Eprint {http://arxiv.org/abs/1707.03419} {arXiv:1707.03419
  [astro-ph.CO]} \BibitemShut {NoStop}%
%%CITATION = ARXIV:1707.03419;%%
\bibitem [{\citenamefont {Garrison-Kimmel}\ \emph {et~al.}(2013)\citenamefont
  {Garrison-Kimmel}, \citenamefont {Rocha}, \citenamefont {Boylan-Kolchin},
  \citenamefont {Bullock},\ and\ \citenamefont
  {Lally}}]{GarrisonKimmel:2013aq}%
  \BibitemOpen
  \bibfield  {author} {\bibinfo {author} {\bibfnamefont {S.}~\bibnamefont
  {Garrison-Kimmel}}, \bibinfo {author} {\bibfnamefont {M.}~\bibnamefont
  {Rocha}}, \bibinfo {author} {\bibfnamefont {M.}~\bibnamefont
  {Boylan-Kolchin}}, \bibinfo {author} {\bibfnamefont {J.}~\bibnamefont
  {Bullock}}, \ and\ \bibinfo {author} {\bibfnamefont {J.}~\bibnamefont
  {Lally}},\ }\href {\doibase 10.1093/mnras/stt984} {\bibfield  {journal}
  {\bibinfo  {journal} {Mon. Not. Roy. Astron. Soc.}\ }\textbf {\bibinfo
  {volume} {433}},\ \bibinfo {pages} {3539} (\bibinfo {year} {2013})},\ \Eprint
  {http://arxiv.org/abs/1301.3137} {arXiv:1301.3137 [astro-ph.CO]} \BibitemShut
  {NoStop}%
%%CITATION = ARXIV:1301.3137;%%
\bibitem [{\citenamefont {Weinberg}\ \emph {et~al.}(2013)\citenamefont
  {Weinberg}, \citenamefont {Bullock}, \citenamefont {Governato}, \citenamefont
  {Kuzio~de Naray},\ and\ \citenamefont {Peter}}]{Weinberg:2013aya}%
  \BibitemOpen
  \bibfield  {author} {\bibinfo {author} {\bibfnamefont {D.~H.}\ \bibnamefont
  {Weinberg}}, \bibinfo {author} {\bibfnamefont {J.~S.}\ \bibnamefont
  {Bullock}}, \bibinfo {author} {\bibfnamefont {F.}~\bibnamefont {Governato}},
  \bibinfo {author} {\bibfnamefont {R.}~\bibnamefont {Kuzio~de Naray}}, \ and\
  \bibinfo {author} {\bibfnamefont {A.~H.~G.}\ \bibnamefont {Peter}},\ }in\
  \href {http://inspirehep.net/record/1237028/files/arXiv:1306.0913.pdf} {\emph
  {\bibinfo {booktitle} {{Sackler Colloquium: Dark Matter Universe: On the
  Threshhold of Discovery Irvine, USA, October 18-20, 2012}}}}\ (\bibinfo
  {year} {2013})\ \Eprint {http://arxiv.org/abs/1306.0913} {arXiv:1306.0913
  [astro-ph.CO]} \BibitemShut {NoStop}%
%%CITATION = ARXIV:1306.0913;%%
\bibitem [{\citenamefont {Pontzen}\ and\ \citenamefont
  {Governato}(2012)}]{Pontzen:2011ty}%
  \BibitemOpen
  \bibfield  {author} {\bibinfo {author} {\bibfnamefont {A.}~\bibnamefont
  {Pontzen}}\ and\ \bibinfo {author} {\bibfnamefont {F.}~\bibnamefont
  {Governato}},\ }\href {\doibase 10.1111/j.1365-2966.2012.20571.x} {\bibfield
  {journal} {\bibinfo  {journal} {Mon. Not. Roy. Astron. Soc.}\ }\textbf
  {\bibinfo {volume} {421}},\ \bibinfo {pages} {3464} (\bibinfo {year}
  {2012})},\ \Eprint {http://arxiv.org/abs/1106.0499} {arXiv:1106.0499
  [astro-ph.CO]} \BibitemShut {NoStop}%
%%CITATION = ARXIV:1106.0499;%%
\bibitem [{\citenamefont {Zolotov}\ \emph {et~al.}(2012)\citenamefont
  {Zolotov}, \citenamefont {Brooks}, \citenamefont {Willman}, \citenamefont
  {Governato}, \citenamefont {Pontzen}, \citenamefont {Christensen},
  \citenamefont {Dekel}, \citenamefont {Quinn}, \citenamefont {Shen},\ and\
  \citenamefont {Wadsley}}]{Zolotov:2012xd}%
  \BibitemOpen
  \bibfield  {author} {\bibinfo {author} {\bibfnamefont {A.}~\bibnamefont
  {Zolotov}}, \bibinfo {author} {\bibfnamefont {A.~M.}\ \bibnamefont {Brooks}},
  \bibinfo {author} {\bibfnamefont {B.}~\bibnamefont {Willman}}, \bibinfo
  {author} {\bibfnamefont {F.}~\bibnamefont {Governato}}, \bibinfo {author}
  {\bibfnamefont {A.}~\bibnamefont {Pontzen}}, \bibinfo {author} {\bibfnamefont
  {C.}~\bibnamefont {Christensen}}, \bibinfo {author} {\bibfnamefont
  {A.}~\bibnamefont {Dekel}}, \bibinfo {author} {\bibfnamefont
  {T.}~\bibnamefont {Quinn}}, \bibinfo {author} {\bibfnamefont
  {S.}~\bibnamefont {Shen}}, \ and\ \bibinfo {author} {\bibfnamefont
  {J.}~\bibnamefont {Wadsley}},\ }\href {\doibase 10.1088/0004-637X/761/1/71}
  {\bibfield  {journal} {\bibinfo  {journal} {Astrophys. J.}\ }\textbf
  {\bibinfo {volume} {761}},\ \bibinfo {pages} {71} (\bibinfo {year} {2012})},\
  \Eprint {http://arxiv.org/abs/1207.0007} {arXiv:1207.0007 [astro-ph.CO]}
  \BibitemShut {NoStop}%
%%CITATION = ARXIV:1207.0007;%%
\bibitem [{\citenamefont {Governato}\ \emph {et~al.}(2012)\citenamefont
  {Governato}, \citenamefont {Zolotov}, \citenamefont {Pontzen}, \citenamefont
  {Christensen}, \citenamefont {Oh}, \citenamefont {Brooks}, \citenamefont
  {Quinn}, \citenamefont {Shen},\ and\ \citenamefont
  {Wadsley}}]{Governato:2012fa}%
  \BibitemOpen
  \bibfield  {author} {\bibinfo {author} {\bibfnamefont {F.}~\bibnamefont
  {Governato}}, \bibinfo {author} {\bibfnamefont {A.}~\bibnamefont {Zolotov}},
  \bibinfo {author} {\bibfnamefont {A.}~\bibnamefont {Pontzen}}, \bibinfo
  {author} {\bibfnamefont {C.}~\bibnamefont {Christensen}}, \bibinfo {author}
  {\bibfnamefont {S.~H.}\ \bibnamefont {Oh}}, \bibinfo {author} {\bibfnamefont
  {A.~M.}\ \bibnamefont {Brooks}}, \bibinfo {author} {\bibfnamefont
  {T.}~\bibnamefont {Quinn}}, \bibinfo {author} {\bibfnamefont
  {S.}~\bibnamefont {Shen}}, \ and\ \bibinfo {author} {\bibfnamefont
  {J.}~\bibnamefont {Wadsley}},\ }\href {\doibase
  10.1111/j.1365-2966.2012.20696.x} {\bibfield  {journal} {\bibinfo  {journal}
  {Mon. Not. Roy. Astron. Soc.}\ }\textbf {\bibinfo {volume} {422}},\ \bibinfo
  {pages} {1231} (\bibinfo {year} {2012})},\ \Eprint
  {http://arxiv.org/abs/1202.0554} {arXiv:1202.0554 [astro-ph.CO]} \BibitemShut
  {NoStop}%
%%CITATION = ARXIV:1202.0554;%%
\bibitem [{\citenamefont {Teyssier}\ \emph {et~al.}(2013)\citenamefont
  {Teyssier}, \citenamefont {Pontzen}, \citenamefont {Dubois},\ and\
  \citenamefont {Read}}]{Teyssier:2012ie}%
  \BibitemOpen
  \bibfield  {author} {\bibinfo {author} {\bibfnamefont {R.}~\bibnamefont
  {Teyssier}}, \bibinfo {author} {\bibfnamefont {A.}~\bibnamefont {Pontzen}},
  \bibinfo {author} {\bibfnamefont {Y.}~\bibnamefont {Dubois}}, \ and\ \bibinfo
  {author} {\bibfnamefont {J.}~\bibnamefont {Read}},\ }\href {\doibase
  10.1093/mnras/sts563} {\bibfield  {journal} {\bibinfo  {journal} {Mon. Not.
  Roy. Astron. Soc.}\ }\textbf {\bibinfo {volume} {429}},\ \bibinfo {pages}
  {3068} (\bibinfo {year} {2013})},\ \Eprint {http://arxiv.org/abs/1206.4895}
  {arXiv:1206.4895 [astro-ph.CO]} \BibitemShut {NoStop}%
%%CITATION = ARXIV:1206.4895;%%
\bibitem [{\citenamefont {Brooks}\ and\ \citenamefont
  {Zolotov}(2014)}]{Brooks:2012vi}%
  \BibitemOpen
  \bibfield  {author} {\bibinfo {author} {\bibfnamefont {A.~M.}\ \bibnamefont
  {Brooks}}\ and\ \bibinfo {author} {\bibfnamefont {A.}~\bibnamefont
  {Zolotov}},\ }\href {\doibase 10.1088/0004-637X/786/2/87} {\bibfield
  {journal} {\bibinfo  {journal} {Astrophys. J.}\ }\textbf {\bibinfo {volume}
  {786}},\ \bibinfo {pages} {87} (\bibinfo {year} {2014})},\ \Eprint
  {http://arxiv.org/abs/1207.2468} {arXiv:1207.2468 [astro-ph.CO]} \BibitemShut
  {NoStop}%
%%CITATION = ARXIV:1207.2468;%%
\bibitem [{\citenamefont {Brooks}\ \emph {et~al.}(2013)\citenamefont {Brooks},
  \citenamefont {Kuhlen}, \citenamefont {Zolotov},\ and\ \citenamefont
  {Hooper}}]{Brooks:2012ah}%
  \BibitemOpen
  \bibfield  {author} {\bibinfo {author} {\bibfnamefont {A.~M.}\ \bibnamefont
  {Brooks}}, \bibinfo {author} {\bibfnamefont {M.}~\bibnamefont {Kuhlen}},
  \bibinfo {author} {\bibfnamefont {A.}~\bibnamefont {Zolotov}}, \ and\
  \bibinfo {author} {\bibfnamefont {D.}~\bibnamefont {Hooper}},\ }\href
  {\doibase 10.1088/0004-637X/765/1/22} {\bibfield  {journal} {\bibinfo
  {journal} {Astrophys. J.}\ }\textbf {\bibinfo {volume} {765}},\ \bibinfo
  {pages} {22} (\bibinfo {year} {2013})},\ \Eprint
  {http://arxiv.org/abs/1209.5394} {arXiv:1209.5394 [astro-ph.CO]} \BibitemShut
  {NoStop}%
%%CITATION = ARXIV:1209.5394;%%
\bibitem [{\citenamefont {Arraki}\ \emph {et~al.}(2014)\citenamefont {Arraki},
  \citenamefont {Klypin}, \citenamefont {More},\ and\ \citenamefont
  {Trujillo-Gomez}}]{Arraki:2012bu}%
  \BibitemOpen
  \bibfield  {author} {\bibinfo {author} {\bibfnamefont {K.~S.}\ \bibnamefont
  {Arraki}}, \bibinfo {author} {\bibfnamefont {A.}~\bibnamefont {Klypin}},
  \bibinfo {author} {\bibfnamefont {S.}~\bibnamefont {More}}, \ and\ \bibinfo
  {author} {\bibfnamefont {S.}~\bibnamefont {Trujillo-Gomez}},\ }\href
  {\doibase 10.1093/mnras/stt2279} {\bibfield  {journal} {\bibinfo  {journal}
  {Mon. Not. Roy. Astron. Soc.}\ }\textbf {\bibinfo {volume} {438}},\ \bibinfo
  {pages} {1466} (\bibinfo {year} {2014})},\ \Eprint
  {http://arxiv.org/abs/1212.6651} {arXiv:1212.6651 [astro-ph.CO]} \BibitemShut
  {NoStop}%
%%CITATION = ARXIV:1212.6651;%%
\bibitem [{\citenamefont {Amorisco}\ \emph {et~al.}(2014)\citenamefont
  {Amorisco}, \citenamefont {Zavala},\ and\ \citenamefont
  {de~Boer}}]{Amorisco:2013uwa}%
  \BibitemOpen
  \bibfield  {author} {\bibinfo {author} {\bibfnamefont {N.~C.}\ \bibnamefont
  {Amorisco}}, \bibinfo {author} {\bibfnamefont {J.}~\bibnamefont {Zavala}}, \
  and\ \bibinfo {author} {\bibfnamefont {T.~J.~L.}\ \bibnamefont {de~Boer}},\
  }\href {\doibase 10.1088/2041-8205/782/2/L39} {\bibfield  {journal} {\bibinfo
   {journal} {Astrophys. J.}\ }\textbf {\bibinfo {volume} {782}},\ \bibinfo
  {pages} {L39} (\bibinfo {year} {2014})},\ \Eprint
  {http://arxiv.org/abs/1309.5958} {arXiv:1309.5958 [astro-ph.CO]} \BibitemShut
  {NoStop}%
%%CITATION = ARXIV:1309.5958;%%
\bibitem [{\citenamefont {Madau}\ \emph {et~al.}(2008)\citenamefont {Madau},
  \citenamefont {Diemand},\ and\ \citenamefont {Kuhlen}}]{Madau:2008fr}%
  \BibitemOpen
  \bibfield  {author} {\bibinfo {author} {\bibfnamefont {P.}~\bibnamefont
  {Madau}}, \bibinfo {author} {\bibfnamefont {J.}~\bibnamefont {Diemand}}, \
  and\ \bibinfo {author} {\bibfnamefont {M.}~\bibnamefont {Kuhlen}},\ }\href
  {\doibase 10.1086/587545} {\bibfield  {journal} {\bibinfo  {journal}
  {Astrophys. J.}\ }\textbf {\bibinfo {volume} {679}},\ \bibinfo {pages} {1260}
  (\bibinfo {year} {2008})},\ \Eprint {http://arxiv.org/abs/0802.2265}
  {arXiv:0802.2265 [astro-ph]} \BibitemShut {NoStop}%
%%CITATION = ARXIV:0802.2265;%%
\bibitem [{\citenamefont {{Angulo}}\ and\ \citenamefont
  {{White}}(2010)}]{2010MNRAS.401.1796A}%
  \BibitemOpen
  \bibfield  {author} {\bibinfo {author} {\bibfnamefont {R.~E.}\ \bibnamefont
  {{Angulo}}}\ and\ \bibinfo {author} {\bibfnamefont {S.~D.~M.}\ \bibnamefont
  {{White}}},\ }\href {\doibase 10.1111/j.1365-2966.2009.15742.x} {\bibfield
  {journal} {\bibinfo  {journal} {Mon. Not. Roy. Astron. Soc.}\ }\textbf
  {\bibinfo {volume} {401}},\ \bibinfo {pages} {1796} (\bibinfo {year}
  {2010})},\ \Eprint {http://arxiv.org/abs/0906.1730} {arXiv:0906.1730}
  \BibitemShut {NoStop}%
\bibitem [{\citenamefont {Rashkov}\ \emph {et~al.}(2012)\citenamefont
  {Rashkov}, \citenamefont {Madau}, \citenamefont {Kuhlen},\ and\ \citenamefont
  {Diemand}}]{Rashkov:2011cq}%
  \BibitemOpen
  \bibfield  {author} {\bibinfo {author} {\bibfnamefont {V.}~\bibnamefont
  {Rashkov}}, \bibinfo {author} {\bibfnamefont {P.}~\bibnamefont {Madau}},
  \bibinfo {author} {\bibfnamefont {M.}~\bibnamefont {Kuhlen}}, \ and\ \bibinfo
  {author} {\bibfnamefont {J.}~\bibnamefont {Diemand}},\ }\href {\doibase
  10.1088/0004-637X/745/2/142} {\bibfield  {journal} {\bibinfo  {journal}
  {Astrophys. J.}\ }\textbf {\bibinfo {volume} {745}},\ \bibinfo {pages} {142}
  (\bibinfo {year} {2012})},\ \Eprint {http://arxiv.org/abs/1106.5583}
  {arXiv:1106.5583 [astro-ph.GA]} \BibitemShut {NoStop}%
%%CITATION = ARXIV:1106.5583;%%
\bibitem [{\citenamefont {Brandt}(2016)}]{Brandt:2016aco}%
  \BibitemOpen
  \bibfield  {author} {\bibinfo {author} {\bibfnamefont {T.~D.}\ \bibnamefont
  {Brandt}},\ }\href {\doibase 10.3847/2041-8205/824/2/L31} {\bibfield
  {journal} {\bibinfo  {journal} {Astrophys. J.}\ }\textbf {\bibinfo {volume}
  {824}},\ \bibinfo {pages} {L31} (\bibinfo {year} {2016})},\ \Eprint
  {http://arxiv.org/abs/1605.03665} {arXiv:1605.03665 [astro-ph.GA]}
  \BibitemShut {NoStop}%
%%CITATION = ARXIV:1605.03665;%%
\bibitem [{\citenamefont {Carr}\ and\ \citenamefont
  {Sakellariadou}(1999)}]{Carr:1997cn}%
  \BibitemOpen
  \bibfield  {author} {\bibinfo {author} {\bibfnamefont {B.~J.}\ \bibnamefont
  {Carr}}\ and\ \bibinfo {author} {\bibfnamefont {M.}~\bibnamefont
  {Sakellariadou}},\ }\href {\doibase 10.1086/307071} {\bibfield  {journal}
  {\bibinfo  {journal} {Astrophys. J.}\ }\textbf {\bibinfo {volume} {516}},\
  \bibinfo {pages} {195} (\bibinfo {year} {1999})}\BibitemShut {NoStop}%
%%CITATION = ASJOA,516,195;%%
\bibitem [{\citenamefont {Wilkinson}\ \emph {et~al.}(2001)\citenamefont
  {Wilkinson}, \citenamefont {Henstock}, \citenamefont {Browne}, \citenamefont
  {Polatidis}, \citenamefont {Augusto}, \citenamefont {Readhead}, \citenamefont
  {Pearson}, \citenamefont {Xu}, \citenamefont {Taylor},\ and\ \citenamefont
  {Vermeulen}}]{Wilkinson:2001vv}%
  \BibitemOpen
  \bibfield  {author} {\bibinfo {author} {\bibfnamefont {P.~N.}\ \bibnamefont
  {Wilkinson}}, \bibinfo {author} {\bibfnamefont {D.~R.}\ \bibnamefont
  {Henstock}}, \bibinfo {author} {\bibfnamefont {I.~W.~A.}\ \bibnamefont
  {Browne}}, \bibinfo {author} {\bibfnamefont {A.~G.}\ \bibnamefont
  {Polatidis}}, \bibinfo {author} {\bibfnamefont {P.}~\bibnamefont {Augusto}},
  \bibinfo {author} {\bibfnamefont {A.~C.~S.}\ \bibnamefont {Readhead}},
  \bibinfo {author} {\bibfnamefont {T.~J.}\ \bibnamefont {Pearson}}, \bibinfo
  {author} {\bibfnamefont {W.}~\bibnamefont {Xu}}, \bibinfo {author}
  {\bibfnamefont {G.~B.}\ \bibnamefont {Taylor}}, \ and\ \bibinfo {author}
  {\bibfnamefont {R.~C.}\ \bibnamefont {Vermeulen}},\ }\href {\doibase
  10.1103/PhysRevLett.86.584} {\bibfield  {journal} {\bibinfo  {journal} {Phys.
  Rev. Lett.}\ }\textbf {\bibinfo {volume} {86}},\ \bibinfo {pages} {584}
  (\bibinfo {year} {2001})},\ \Eprint {http://arxiv.org/abs/astro-ph/0101328}
  {arXiv:astro-ph/0101328 [astro-ph]} \BibitemShut {NoStop}%
%%CITATION = ASTRO-PH/0101328;%%
\bibitem [{\citenamefont {Carr}\ \emph {et~al.}(2016)\citenamefont {Carr},
  \citenamefont {Kuhnel},\ and\ \citenamefont {Sandstad}}]{Carr:2016drx}%
  \BibitemOpen
  \bibfield  {author} {\bibinfo {author} {\bibfnamefont {B.}~\bibnamefont
  {Carr}}, \bibinfo {author} {\bibfnamefont {F.}~\bibnamefont {Kuhnel}}, \ and\
  \bibinfo {author} {\bibfnamefont {M.}~\bibnamefont {Sandstad}},\ }\href
  {\doibase 10.1103/PhysRevD.94.083504} {\bibfield  {journal} {\bibinfo
  {journal} {Phys. Rev.}\ }\textbf {\bibinfo {volume} {D94}},\ \bibinfo {pages}
  {083504} (\bibinfo {year} {2016})},\ \Eprint
  {http://arxiv.org/abs/1607.06077} {arXiv:1607.06077 [astro-ph.CO]}
  \BibitemShut {NoStop}%
%%CITATION = ARXIV:1607.06077;%%
\bibitem [{\citenamefont {Carr}\ \emph {et~al.}(2017)\citenamefont {Carr},
  \citenamefont {Raidal}, \citenamefont {Tenkanen}, \citenamefont {Vaskonen},\
  and\ \citenamefont {Veerm{\"a}e}}]{Carr:2017jsz}%
  \BibitemOpen
  \bibfield  {author} {\bibinfo {author} {\bibfnamefont {B.}~\bibnamefont
  {Carr}}, \bibinfo {author} {\bibfnamefont {M.}~\bibnamefont {Raidal}},
  \bibinfo {author} {\bibfnamefont {T.}~\bibnamefont {Tenkanen}}, \bibinfo
  {author} {\bibfnamefont {V.}~\bibnamefont {Vaskonen}}, \ and\ \bibinfo
  {author} {\bibfnamefont {H.}~\bibnamefont {Veerm{\"a}e}},\ }\href {\doibase
  10.1103/PhysRevD.96.023514} {\bibfield  {journal} {\bibinfo  {journal} {Phys.
  Rev.}\ }\textbf {\bibinfo {volume} {D96}},\ \bibinfo {pages} {023514}
  (\bibinfo {year} {2017})},\ \Eprint {http://arxiv.org/abs/1705.05567}
  {arXiv:1705.05567 [astro-ph.CO]} \BibitemShut {NoStop}%
%%CITATION = ARXIV:1705.05567;%%
\bibitem [{\citenamefont {Poulin}\ \emph {et~al.}(2017)\citenamefont {Poulin},
  \citenamefont {Serpico}, \citenamefont {Calore}, \citenamefont {Clesse},\
  and\ \citenamefont {Kohri}}]{Poulin:2017bwe}%
  \BibitemOpen
  \bibfield  {author} {\bibinfo {author} {\bibfnamefont {V.}~\bibnamefont
  {Poulin}}, \bibinfo {author} {\bibfnamefont {P.~D.}\ \bibnamefont {Serpico}},
  \bibinfo {author} {\bibfnamefont {F.}~\bibnamefont {Calore}}, \bibinfo
  {author} {\bibfnamefont {S.}~\bibnamefont {Clesse}}, \ and\ \bibinfo {author}
  {\bibfnamefont {K.}~\bibnamefont {Kohri}},\ }\href@noop {} {\  (\bibinfo
  {year} {2017})},\ \Eprint {http://arxiv.org/abs/1707.04206} {arXiv:1707.04206
  [astro-ph.CO]} \BibitemShut {NoStop}%
%%CITATION = ARXIV:1707.04206;%%
\bibitem [{\citenamefont {Cyr-Racine}\ \emph {et~al.}(2016)\citenamefont
  {Cyr-Racine}, \citenamefont {Moustakas}, \citenamefont {Keeton},
  \citenamefont {Sigurdson},\ and\ \citenamefont
  {Gilman}}]{Cyr-Racine:2015jwa}%
  \BibitemOpen
  \bibfield  {author} {\bibinfo {author} {\bibfnamefont {F.-Y.}\ \bibnamefont
  {Cyr-Racine}}, \bibinfo {author} {\bibfnamefont {L.~A.}\ \bibnamefont
  {Moustakas}}, \bibinfo {author} {\bibfnamefont {C.~R.}\ \bibnamefont
  {Keeton}}, \bibinfo {author} {\bibfnamefont {K.}~\bibnamefont {Sigurdson}}, \
  and\ \bibinfo {author} {\bibfnamefont {D.~A.}\ \bibnamefont {Gilman}},\
  }\href {\doibase 10.1103/PhysRevD.94.043505} {\bibfield  {journal} {\bibinfo
  {journal} {Phys. Rev.}\ }\textbf {\bibinfo {volume} {D94}},\ \bibinfo {pages}
  {043505} (\bibinfo {year} {2016})},\ \Eprint
  {http://arxiv.org/abs/1506.01724} {arXiv:1506.01724 [astro-ph.CO]}
  \BibitemShut {NoStop}%
%%CITATION = ARXIV:1506.01724;%%
\bibitem [{\citenamefont {Ibata}\ \emph {et~al.}(2001)\citenamefont {Ibata},
  \citenamefont {Lewis}, \citenamefont {Irwin}, \citenamefont {Totten},\ and\
  \citenamefont {Quinn}}]{Ibata:2000pu}%
  \BibitemOpen
  \bibfield  {author} {\bibinfo {author} {\bibfnamefont {R.}~\bibnamefont
  {Ibata}}, \bibinfo {author} {\bibfnamefont {G.~F.}\ \bibnamefont {Lewis}},
  \bibinfo {author} {\bibfnamefont {M.}~\bibnamefont {Irwin}}, \bibinfo
  {author} {\bibfnamefont {E.}~\bibnamefont {Totten}}, \ and\ \bibinfo {author}
  {\bibfnamefont {T.~R.}\ \bibnamefont {Quinn}},\ }\href {\doibase
  10.1086/320060} {\bibfield  {journal} {\bibinfo  {journal} {Astrophys. J.}\
  }\textbf {\bibinfo {volume} {551}},\ \bibinfo {pages} {294} (\bibinfo {year}
  {2001})},\ \Eprint {http://arxiv.org/abs/astro-ph/0004011}
  {arXiv:astro-ph/0004011 [astro-ph]} \BibitemShut {NoStop}%
%%CITATION = ASTRO-PH/0004011;%%
\bibitem [{\citenamefont {Johnston}\ \emph {et~al.}(2002)\citenamefont
  {Johnston}, \citenamefont {Spergel},\ and\ \citenamefont
  {Haydn}}]{Johnston:2001wh}%
  \BibitemOpen
  \bibfield  {author} {\bibinfo {author} {\bibfnamefont {K.~V.}\ \bibnamefont
  {Johnston}}, \bibinfo {author} {\bibfnamefont {D.~N.}\ \bibnamefont
  {Spergel}}, \ and\ \bibinfo {author} {\bibfnamefont {C.}~\bibnamefont
  {Haydn}},\ }\href {\doibase 10.1086/339791} {\bibfield  {journal} {\bibinfo
  {journal} {Astrophys. J.}\ }\textbf {\bibinfo {volume} {570}},\ \bibinfo
  {pages} {656} (\bibinfo {year} {2002})},\ \Eprint
  {http://arxiv.org/abs/astro-ph/0111196} {arXiv:astro-ph/0111196 [astro-ph]}
  \BibitemShut {NoStop}%
%%CITATION = ASTRO-PH/0111196;%%
\bibitem [{\citenamefont {Erkal}\ and\ \citenamefont
  {Belokurov}(2015)}]{Erkal:2015kqa}%
  \BibitemOpen
  \bibfield  {author} {\bibinfo {author} {\bibfnamefont {D.}~\bibnamefont
  {Erkal}}\ and\ \bibinfo {author} {\bibfnamefont {V.}~\bibnamefont
  {Belokurov}},\ }\href {\doibase 10.1093/mnras/stv2122} {\bibfield  {journal}
  {\bibinfo  {journal} {Mon. Not. Roy. Astron. Soc.}\ }\textbf {\bibinfo
  {volume} {454}},\ \bibinfo {pages} {3542} (\bibinfo {year} {2015})},\ \Eprint
  {http://arxiv.org/abs/1507.05625} {arXiv:1507.05625 [astro-ph.GA]}
  \BibitemShut {NoStop}%
%%CITATION = ARXIV:1507.05625;%%
\bibitem [{\citenamefont {Bovy}\ \emph {et~al.}(2016)\citenamefont {Bovy},
  \citenamefont {Bahmanyar}, \citenamefont {Fritz},\ and\ \citenamefont
  {Kallivayalil}}]{Bovy:2016chl}%
  \BibitemOpen
  \bibfield  {author} {\bibinfo {author} {\bibfnamefont {J.}~\bibnamefont
  {Bovy}}, \bibinfo {author} {\bibfnamefont {A.}~\bibnamefont {Bahmanyar}},
  \bibinfo {author} {\bibfnamefont {T.~K.}\ \bibnamefont {Fritz}}, \ and\
  \bibinfo {author} {\bibfnamefont {N.}~\bibnamefont {Kallivayalil}},\ }\href
  {\doibase 10.3847/1538-4357/833/1/31} {\bibfield  {journal} {\bibinfo
  {journal} {Astrophys. J.}\ }\textbf {\bibinfo {volume} {833}},\ \bibinfo
  {pages} {31} (\bibinfo {year} {2016})},\ \Eprint
  {http://arxiv.org/abs/1609.01298} {arXiv:1609.01298 [astro-ph.GA]}
  \BibitemShut {NoStop}%
%%CITATION = ARXIV:1609.01298;%%
\bibitem [{\citenamefont {{Kruijssen}}(2014)}]{2014CQGra..31x4006K}%
  \BibitemOpen
  \bibfield  {author} {\bibinfo {author} {\bibfnamefont {J.~M.~D.}\
  \bibnamefont {{Kruijssen}}},\ }\href {\doibase
  10.1088/0264-9381/31/24/244006} {\bibfield  {journal} {\bibinfo  {journal}
  {Classical and Quantum Gravity}\ }\textbf {\bibinfo {volume} {31}},\ \bibinfo
  {eid} {244006} (\bibinfo {year} {2014})},\ \Eprint
  {http://arxiv.org/abs/1407.2953} {arXiv:1407.2953} \BibitemShut {NoStop}%
\bibitem [{\citenamefont {Adam}\ \emph {et~al.}(2016)\citenamefont {Adam} \emph
  {et~al.}}]{Adam:2016hgk}%
  \BibitemOpen
  \bibfield  {author} {\bibinfo {author} {\bibfnamefont {R.}~\bibnamefont
  {Adam}} \emph {et~al.} (\bibinfo {collaboration} {Planck}),\ }\href {\doibase
  10.1051/0004-6361/201628897} {\bibfield  {journal} {\bibinfo  {journal}
  {Astron. Astrophys.}\ }\textbf {\bibinfo {volume} {596}},\ \bibinfo {pages}
  {A108} (\bibinfo {year} {2016})},\ \Eprint {http://arxiv.org/abs/1605.03507}
  {arXiv:1605.03507 [astro-ph.CO]} \BibitemShut {NoStop}%
%%CITATION = ARXIV:1605.03507;%%
\end{thebibliography}%

\end{document}